\documentclass[journal]{IEEEtran}
\usepackage[noadjust]{cite}
\ifCLASSINFOpdf
	\usepackage[pdftex]{graphicx}
    \DeclareGraphicsExtensions{.pdf,.png}
\else
	\usepackage[dvips]{graphicx}
    \DeclareGraphicsExtensions{.pdf,.png}
\fi
\usepackage{amsmath,amssymb,mathtools}
\usepackage{parskip}
\usepackage{tikz}
\usetikzlibrary{calc}
\newcommand{\E}{\operatorname{E}}
\DeclareMathOperator*{\minimize}{\text{minimize}}
\begin{document}
\title{A Hardware Realization of Superresolution Combining Random Coding and Blurring}
\author{Kevin~Beale,~\IEEEmembership{Student~Member,~IEEE,}
	Jianbo~Chen,~\IEEEmembership{Student~Member,~IEEE,}
	Kevin~Kelly,~\IEEEmembership{Member,~IEEE,}
	and~Justin~Romberg,~\IEEEmembership{Fellow,~IEEE}%
\thanks{KB and JC contributed equally to this work.}%
\thanks{KB and JR are affiliated with the Department of Electrical and Computer Engineering, Georgia Institute of Technology, Atlanta, GA, 30332 USA.}%
\thanks{JC and KK are affiliated with the Department of Electrical and Computer Engineering, Rice University, Houston, TX, 77251, USA.}%
\thanks{KB and JR were supported by a grant from the Packard Foundation. KK and JC gratefully acknowledge grants from the NSF (CHE-1610453) and the AFOSR (FA8651-16-C-0185) in support of this research.}%
\thanks{Author emails: KB (kdbeale@gmail.com), JC (chenjianbo719@gmail.com), KK (kkelly@rice.edu), JR (jrom@ece.gatech.edu).}%
\thanks{Manuscript received September 20, 2018.}
}
\markboth{IEEE Transactions on Computational Imaging,~Vol.~?, No.~?, Month~2018}
{\MakeLowercase{\textit{Beale et al.}}: Hardware Realization of Superresolution}
\maketitle
\begin{abstract}
Resolution enhancements are often desired in imaging applications where high-resolution sensor arrays are difficult to obtain. Many computational imaging methods have been proposed to encode high-resolution scene information on low-resolution sensors by cleverly modulating light from the scene before it hits the sensor. These methods often require movement of some portion of the imaging apparatus or only acquire images up to the resolution of a modulating element. Here a technique is presented for resolving beyond the resolutions of both a pointwise-modulating mask element and a sensor array through the introduction of a controlled blur into the optical pathway. The analysis contains an intuitive and exact expression for the overall superresolvability of the system, and arguments are presented to explain how the combination of random coding and blurring makes the superresolution problem well-posed. Experimental results demonstrate that a resolution enhancement of approximately $4\times$ is possible in practice using standard optical components, without mechanical motion of the imaging apparatus, and without any a priori assumptions on scene structure.
\end{abstract}
\begin{IEEEkeywords}
Superresolution, Random Coding, Digital Micromirror Device
\end{IEEEkeywords}
\section{Introduction} 
\IEEEPARstart{F}{or} most of the history of imaging, the use of photographic film (i.e. analog sensors) meant that the resolution of images was largely a function of how well the impulse response of the collection optics approximated the delta function at the image plane. With the advent of digital sensor arrays, the outputs of which could more easily be processed computationally, new possibilities emerged for sensing and interpreting images. The field of computational imaging is now concerned with the co-design of light-modulation schemes and image reconstruction algorithms to better capture desired scene information. This typically leads to imaging systems which capture measurements that do not resemble traditional photographs, but contain scene information in an encoded form that may be recovered with suitable algorithms.

Adding a programmable mask to an imaging system is a simple modification that can significantly increase its flexibility  \cite{ProgrammableImaging}.  In applications where high-resolution sensor arrays are unavailable or prohibitively expensive, adding a programmable mask can improve the resolution of a low-resolution focal plane array, even those consisting of a single sensor \cite{SinglePixelCamera}. In a configuration where the mask pointwise-attenuates the scene, a series of modulated images may be measured on the low-resolution array, and a high-resolution image computed from these measurements. Measuring a set of images using a sufficient number of diverse (or even random) mask patterns allows us to reconstruct the scene up to the resolution of the programmable mask.

The main contribution of this paper is to show that if an imaging system using a pointwise-modulating programmable mask is followed by an appropriate blurring, the image can be \emph{superresolved} past the resolutions of both the mask and sensor array.

\subsection{Superresolution}
The resolution of a conventional imaging system is typically limited by either the quality of its optical components or the resolution of the sensor array. For systems not already imaging at the diffraction limit, many methods exist for obtaining enhanced resolution images from a series of low-resolution measurements, collectively known as \emph{multiframe superresolution} techniques. In its early manifestations, superresolution consisted of acquiring multiple images from different, barely varying perspectives and merging them into a high-resolution image \cite{SuperresReview}. These techniques interleave the pixels of multiple images and then interpolate to obtain the scene at a higher resolution. However, obtaining samples at the sub-pixel displacements required for meaningful enhancement from a series of randomly varying perspectives is difficult, typically requiring a fairly large number of measurements and a complex registration procedure. Better and more reliable results can be obtained by introducing precisely known, structured modulations into the imaging process. 

Precise modulations can be introduced in the form of exact sub-pixel translations using a motorized stage \cite{PsuedorandomPhaseSuperres}, patterned illumination \cite{ActiveCompImagingSuperres}, a sequence of controlled point spread functions \cite{GeometricSuperresWithDMD}, placing a coded mask in the aperture \cite{CompressiveCodedApSuperres}, random lenses \cite{RandomLensImaging}, and many other techniques. Introducing a high-resolution occlusive programmable mask somewhere into the optical pathway is a simple way of introducing high-resolution variability, and eliminates the need for fine mechanical movements. Combined with reconstruction algorithms that exploit highly efficient image models, computational imaging schemes using occlusive masks are able to acquire images at the resolution of the modulating mask with a small number of total measurements obtained on a low-resolution sensing array.

\subsection{Related Work}
Introducing light modulators and kernels into the optical path to increase the efficiency of the acquisition has a long history in computational imaging.  In this section, we contrast our contributions against previous methods with similar motivations.

In \cite{PsuedorandomPhaseSuperres}, Ashok and Neifeld considered introducing a blurring operation into a multiframe superresolution system based on mechanical shifts, demonstrating that the system with a blur outperformed one that was in-focus. An extended PSF was created by placing a psuedorandom phase mask with optimized roughness and correlation length into the aperture. In contrast to \cite{PsuedorandomPhaseSuperres}, the proposed method grants superresolution capability to a system rather than improves a system that can already superresolve, and uses varying patterns on a programmable mask rather than mechanical shifts to obtain sufficiently diverse information about the scene to perform superresolution. A merger of our approach with this one could be possible by using a psuedorandom phase mask optimized in the fashion \cite{PsuedorandomPhaseSuperres} describes to implement the blur in our system. Such a system would interestingly employ randomness in both the coding and blurring operations.

In a more recent work, Kashter et al. superresolved beyond the diffraction limit in a digital holography configuration by placing a psuedorandom phase mask in front of the aperture \cite{ResolvingImagesByBlurring}. While the phase mask naturally introduces a distortion, in their case it acts to retain information about the higher spatial frequencies that would have been filtered out by the original system's limited numerical aperture. If the point spread function is measured prior to imaging, it can be correlated with the diffused hologram measured in a particular scene to produce a superresolved hologram. Both \cite{ResolvingImagesByBlurring} and \cite{PsuedorandomPhaseSuperres} demonstrate that introducing a controlled blurring operation into the optical pathway can actually be advantageous, provided that the blurring can be stably inverted. The same principle applies to this work as well, although in our case the blur is introduced after a coding stage.

Zlotnik et al. used a combination of two digital micromirror devices (DMDs) in the aperture and intermediate image planes to realize a random operator for compressive sensing, where a DMD in the aperture was used to generate random point spread functions and a DMD in an intermediate image plane was used for pointwise coding \cite{GeometricSuperresWithDMD}. This work addressed the possibility of pointwise modulating the scene with a DMD, but relied on fine mechanical positioning of the DMD relative to the sensor or a second DMD in the aperture to obtain high-resolution information. Rather than requiring two DMDs or perfect alignment, our method involves only the rough positioning of a single lens after the coding stage and only requires a single DMD. Along with this simplified system, we present a more detailed analysis of the expected performance.

In \cite{LightFieldApSuperres}, Mohan et al. merged multiple images differing by sub-pixel shifts by combining a mask in the aperture plane with a slightly defocused lens. With the mask component sizes fixed and an appropriately defocused lens, changing which element is open on the mask shifts a blurred image of the scene by a precise fraction of a pixel. While the idea of introducing a defocus in a system with an occlusive mask to enable superresolution is similar to what we propose here, the function of the mask in our approach is fundamentally different: here the mask point-wise modulates a focused image of the scene and then blurs it, rather than shifting a blurred scene image. It should be noted that \cite{LightFieldApSuperres} found that the blur kernel when the aperture is fully open should be exactly one sensor pixel in diameter, whereas in our approach the desired blur size is 1.5-2 sensor pixels in diameter. Our method could potentially offer several advantages over \cite{LightFieldApSuperres}, based on advantages that a superresolution method using pointwise modulation would have over a method using sub-pixel shifting. For example in applications where deliberately masking out bright elements is desirable to avoid glare, as in certain astronomical imaging scenarios, the proposed method offers a distinct advantage over \cite{LightFieldApSuperres}, where the entire scene is sampled in every measurement. Our method could also be adapted to scenarios where a variable-resolution image is desired, by adapting the set of masks to contain more variability in the regions where high-resolution information is desired, whereas \cite{LightFieldApSuperres} necessarily acquires the full scene at uniform resolution.

Our work also points to the possibility of extending the achievable resolution of mask-based compressive imaging systems in a way that is straightforward and relatively easy to implement. The core idea of compressive sensing \cite{Don06} is to exploit sparsity-based image models to make sensing more efficient, typically using random linear measurements. The prototypical imaging example of compressive sensing is the single-pixel camera \cite{SinglePixelCamera}, which measures a series of inner products between the scene and random patterns displayed on an occlusive mask. Notably, both the single-pixel camera and its straightforward extension to a multi-pixel sensor \cite{CSSuperres} both only reconstruct the scene at the resolution of the mask. 

Instead of using a programmable mask to realize a random measurement operator for compressive sensing, it is possible to use any optical element that introduces sufficient distortions into the optical pathway such that each sensor element effectively measures a random combination of scene intensities. Psuedorandom phase masks have been to perform compressive imaging at both optical \cite{CSOpticalSuperres} and infrared \cite{CSInfraredSuperres} wavelengths. In \cite{CSSuperresImager} it was demonstrated that even simple spherical aberration could be used to create a sufficiently diverse measurement operator. Multiply scattering materials are another way of introducing randomness for performing compressive imaging using coherent light \cite{ImagingWithNature}. However, digital micromirror devices, which are used in the single-pixel camera as well as the proposed method, have a number of advantages over these alternative optical modulators. DMDs are high-resolution, high-speed, high-precision light modulators. They can be adapted to (and operate consistently between) different wavelengths; different mirror coatings can be applied at relatively low-cost. DMDs are also optically efficient, reflecting practically 50\% of the incoming photons in the case of a half-on pattern, whereas phase masks can absorb a larger percentage of the light. Speed-wise, DMDs can be switched very rapidly (in the tens of KHz) relative to liquid crystal-based spatial light modulators. DMDs are clearly more flexible in the patterns that can be displayed than fixed masks, which have various manufacturing constraints. For these reasons, a DMD is an excellent choice of optical modulator for a compressive or computational imaging system. Demonstrating a way to resolve beyond the resolution of the mask likely implies that compressive imaging systems using a DMD to implement a random measurement operator can be similarly extended to resolve beyond both mask and sensor.

\section{Mathematical Model} 

We now present the basic imaging architecture under consideration along with a corresponding mathematical model. After an appropriate discretization, the problem of superresolving amounts to solving a system of linear equations. Our ability to superresolve an arbitrary scene is then determined by the eigenvalue spectrum of the resulting system matrix, giving us a systematic method to compare different choices of blur kernels and the effect of using multiple masks.  In Section~\ref{sec:blurringsuper} below, we show how introducing the blur after the modulation allows this spectrum to be bounded away from zero.

\begin{figure}[h]
\centering
  \tikz{
  \coordinate (A) at (0,0);        
  \coordinate (B) at (2.3,0.7);  
  \coordinate (C) at (3,-0.5);    
  \coordinate (D) at (5.3, 0.2); 
  \coordinate (E) at (6, -1);      
  \node at (A) [above right] {\includegraphics[scale=0.5]{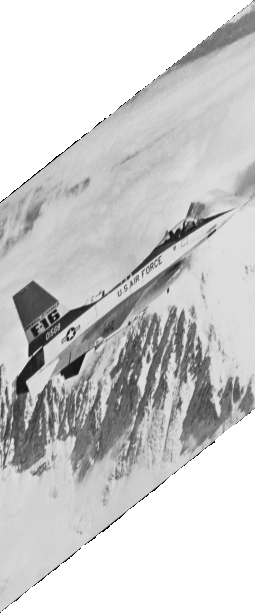}};
  \draw[thick] ($(A) + (0.11,0.12)$) -- ($(A) + (1.48,1.2)$) -- ($(A) + (1.48,3.39)$) -- ($(A) + (0.11,2.3)$) -- cycle;
  \node at ($(A) + (0.6,3.4)$) {Scene};
  \draw (B) to[bend right] ($(B) + (0,1.75)$) node[above] {Lens} to[bend right] (B);
  \draw[thick] ($(C) + (0.11,0.12)$) -- ($(C) + (1.44,1.2)$) -- ($(C) + (1.44,3.29)$) -- ($(C) + (0.11,2.2)$) -- cycle;
  \draw ($(C) + (0.23,0.22)$) -- ($(C) + (0.23,2.3)$);
  \draw ($(C) + (0.33,0.31)$) -- ($(C) + (0.33,2.4)$);
  \draw ($(C) + (0.43,0.39)$) -- ($(C) + (0.43,2.48)$);
  \draw ($(C) + (0.53,0.48)$) -- ($(C) + (0.53,2.55)$);
  \draw ($(C) + (0.63,0.55)$) -- ($(C) + (0.63,2.62)$);
  \draw ($(C) + (0.73,0.64)$) -- ($(C) + (0.73,2.72)$);
  \draw ($(C) + (0.83,0.69)$) -- ($(C) + (0.83,2.8)$);
  \draw ($(C) + (0.93,0.77)$) -- ($(C) + (0.93,2.85)$);
  \draw ($(C) + (1.03,0.85)$) -- ($(C) + (1.03,2.95)$);
  \draw ($(C) + (1.13,0.95)$) -- ($(C) + (1.13,3.05)$);
  \draw ($(C) + (1.23,1.02)$) -- ($(C) + (1.23,3.12)$);
  \draw ($(C) + (1.33,1.1)$) -- ($(C) + (1.33,3.21)$);
  \draw ($(C) + (0.11,0.28)$) -- ($(C) + (1.45,1.37)$);
  \draw ($(C) + (0.11,0.44)$) -- ($(C) + (1.45,1.53)$);
  \draw ($(C) + (0.11,0.6)$) -- ($(C) + (1.45,1.69)$);
  \draw ($(C) + (0.11,0.76)$) -- ($(C) + (1.45,1.85)$);
  \draw ($(C) + (0.11,0.92)$) -- ($(C) + (1.45,2.01)$);
  \draw ($(C) + (0.11,1.08)$) -- ($(C) + (1.45,2.17)$);
  \draw ($(C) + (0.11,1.24)$) -- ($(C) + (1.45,2.33)$);
  \draw ($(C) + (0.11,1.4)$) -- ($(C) + (1.45,2.49)$);
  \draw ($(C) + (0.11,1.56)$) -- ($(C) + (1.45,2.65)$);
  \draw ($(C) + (0.11,1.72)$) -- ($(C) + (1.45,2.81)$);
  \draw ($(C) + (0.11,1.88)$) -- ($(C) + (1.45,2.97)$);
  \draw ($(C) + (0.11,2.04)$) -- ($(C) + (1.45,3.13)$);
  \coordinate (p1) at ($(C) + (0.23,0.22)$);
  \filldraw[fill=black] (p1) -- ($(p1) + (0.1,0.1)$) -- ($(p1) + (0.1,0.23)$) -- ($(p1) + (0,0.15)$) -- cycle;
  \coordinate (p2) at ($(C) + (0.43,0.39)$);
  \filldraw[fill=black] (p2) -- ($(p2) + (0.1,0.1)$) -- ($(p2) + (0.1,0.23)$) -- ($(p2) + (0,0.15)$) -- cycle;
  \coordinate (p3) at ($(C) + (0.23,0.54)$);
  \filldraw[fill=black] (p3) -- ($(p3) + (0.1,0.1)$) -- ($(p3) + (0.1,0.23)$) -- ($(p3) + (0,0.15)$) -- cycle;
  \coordinate (p4) at ($(C) + (0.23,0.86)$);
  \filldraw[fill=black] (p4) -- ($(p4) + (0.1,0.1)$) -- ($(p4) + (0.1,0.23)$) -- ($(p4) + (0,0.15)$) -- cycle;
  \coordinate (p5) at ($(C) + (0.23,1.66)$);
  \filldraw[fill=black] (p5) -- ($(p5) + (0.1,0.1)$) -- ($(p5) + (0.1,0.23)$) -- ($(p5) + (0,0.15)$) -- cycle;
  \coordinate (p6) at ($(C) + (0.23,1.82)$);
  \filldraw[fill=black] (p6) -- ($(p6) + (0.1,0.1)$) -- ($(p6) + (0.1,0.23)$) -- ($(p6) + (0,0.15)$) -- cycle;
  \coordinate (p7) at ($(C) + (0.43,0.86)$);
  \filldraw[fill=black] (p7) -- ($(p7) + (0.1,0.1)$) -- ($(p7) + (0.1,0.23)$) -- ($(p7) + (0,0.15)$) -- cycle;  
  \coordinate (p8) at ($(C) + (0.43,1.34)$);
  \filldraw[fill=black] (p8) -- ($(p8) + (0.1,0.1)$) -- ($(p8) + (0.1,0.23)$) -- ($(p8) + (0,0.15)$) -- cycle;
  \coordinate (p9) at ($(C) + (0.63,0.7)$);
  \filldraw[fill=black] (p9) -- ($(p9) + (0.1,0.1)$) -- ($(p9) + (0.1,0.23)$) -- ($(p9) + (0,0.15)$) -- cycle;
  \coordinate (p10) at ($(C) + (0.63,0.86)$);
  \filldraw[fill=black] (p10) -- ($(p10) + (0.1,0.1)$) -- ($(p10) + (0.1,0.23)$) -- ($(p10) + (0,0.15)$) -- cycle;
  \coordinate (p11) at ($(C) + (0.63,1.98)$);
  \filldraw[fill=black] (p11) -- ($(p11) + (0.1,0.1)$) -- ($(p11) + (0.1,0.23)$) -- ($(p11) + (0,0.15)$) -- cycle;  
  \coordinate (p12) at ($(C) + (0.12,0.61)$);
  \filldraw[fill=black] (p12) -- ($(p12) + (0.1,0.1)$) -- ($(p12) + (0.1,0.23)$) -- ($(p12) + (0,0.15)$) -- cycle;
  \coordinate (p13) at ($(C) + (0.12,1.25)$);
  \filldraw[fill=black] (p13) -- ($(p13) + (0.1,0.1)$) -- ($(p13) + (0.1,0.23)$) -- ($(p13) + (0,0.15)$) -- cycle;
  \coordinate (p14) at ($(C) + (0.12,1.41)$);
  \filldraw[fill=black] (p14) -- ($(p14) + (0.1,0.1)$) -- ($(p14) + (0.1,0.23)$) -- ($(p14) + (0,0.15)$) -- cycle;
  \coordinate (p15) at ($(C) + (0.93,0.78)$);
  \filldraw[fill=black] (p15) -- ($(p15) + (0.1,0.1)$) -- ($(p15) + (0.1,0.23)$) -- ($(p15) + (0,0.15)$) -- cycle;
  \coordinate (p16) at ($(C) + (0.33,0.61)$);
  \filldraw[fill=black] (p16) -- ($(p16) + (0.1,0.1)$) -- ($(p16) + (0.1,0.23)$) -- ($(p16) + (0,0.15)$) -- cycle;
  \coordinate (p17) at ($(C) + (0.33,1.41)$);
  \filldraw[fill=black] (p17) -- ($(p17) + (0.1,0.1)$) -- ($(p17) + (0.1,0.23)$) -- ($(p17) + (0,0.15)$) -- cycle;
  \coordinate (p17) at ($(C) + (0.33,2.21)$);
  \filldraw[fill=black] (p17) -- ($(p17) + (0.1,0.1)$) -- ($(p17) + (0.1,0.23)$) -- ($(p17) + (0,0.15)$) -- cycle;
  \coordinate (p18) at ($(C) + (0.53,1.1)$);
  \filldraw[fill=black] (p18) -- ($(p18) + (0.1,0.1)$) -- ($(p18) + (0.1,0.23)$) -- ($(p18) + (0,0.15)$) -- cycle;
  \coordinate (p19) at ($(C) + (0.53,1.74)$);
  \filldraw[fill=black] (p19) -- ($(p19) + (0.1,0.1)$) -- ($(p19) + (0.1,0.23)$) -- ($(p19) + (0,0.15)$) -- cycle;
  \coordinate (p20) at ($(C) + (0.53,2.06)$);
  \filldraw[fill=black] (p20) -- ($(p20) + (0.1,0.1)$) -- ($(p20) + (0.1,0.23)$) -- ($(p20) + (0,0.15)$) -- cycle;
  \coordinate (p21) at ($(C) + (0.73,2.38)$);
  \filldraw[fill=black] (p21) -- ($(p21) + (0.1,0.1)$) -- ($(p21) + (0.1,0.23)$) -- ($(p21) + (0,0.15)$) -- cycle;
  \coordinate (p22) at ($(C) + (0.73,1.26)$);
  \filldraw[fill=black] (p22) -- ($(p22) + (0.1,0.1)$) -- ($(p22) + (0.1,0.23)$) -- ($(p22) + (0,0.15)$) -- cycle;
  \coordinate (p23) at ($(C) + (0.73,1.1)$);
  \filldraw[fill=black] (p23) -- ($(p23) + (0.1,0.1)$) -- ($(p23) + (0.1,0.23)$) -- ($(p23) + (0,0.15)$) -- cycle;
  \coordinate (p24) at ($(C) + (0.83,1.5)$);
  \filldraw[fill=black] (p24) -- ($(p24) + (0.1,0.1)$) -- ($(p24) + (0.1,0.23)$) -- ($(p24) + (0,0.15)$) -- cycle;
  \coordinate (p25) at ($(C) + (0.93,1.42)$);
  \filldraw[fill=black] (p25) -- ($(p25) + (0.1,0.1)$) -- ($(p25) + (0.1,0.23)$) -- ($(p25) + (0,0.15)$) -- cycle;
  \coordinate (p26) at ($(C) + (0.93,2.38)$);
  \filldraw[fill=black] (p26) -- ($(p26) + (0.1,0.1)$) -- ($(p26) + (0.1,0.23)$) -- ($(p26) + (0,0.15)$) -- cycle;
  \coordinate (p27) at ($(C) + (1.03,2.16)$);
  \filldraw[fill=black] (p27) -- ($(p27) + (0.1,0.1)$) -- ($(p27) + (0.1,0.23)$) -- ($(p27) + (0,0.15)$) -- cycle;
  \coordinate (p28) at ($(C) + (1.03,1.83)$);
  \filldraw[fill=black] (p28) -- ($(p28) + (0.1,0.1)$) -- ($(p28) + (0.1,0.23)$) -- ($(p28) + (0,0.15)$) -- cycle;
  \coordinate (p29) at ($(C) + (1.13,1.75)$);
  \filldraw[fill=black] (p29) -- ($(p29) + (0.1,0.1)$) -- ($(p29) + (0.1,0.23)$) -- ($(p29) + (0,0.15)$) -- cycle;
  \coordinate (p30) at ($(C) + (1.13,1.27)$);
  \filldraw[fill=black] (p30) -- ($(p30) + (0.1,0.1)$) -- ($(p30) + (0.1,0.23)$) -- ($(p30) + (0,0.15)$) -- cycle;
  \coordinate (p31) at ($(C) + (1.23,1.04)$);
  \filldraw[fill=black] (p31) -- ($(p31) + (0.1,0.1)$) -- ($(p31) + (0.1,0.23)$) -- ($(p31) + (0,0.15)$) -- cycle;
  \coordinate (p32) at ($(C) + (1.23,2.47)$);
  \filldraw[fill=black] (p32) -- ($(p32) + (0.1,0.1)$) -- ($(p32) + (0.1,0.23)$) -- ($(p32) + (0,0.15)$) -- cycle;
  \coordinate (p33) at ($(C) + (1.33,1.27)$);
  \filldraw[fill=black] (p33) -- ($(p33) + (0.1,0.1)$) -- ($(p33) + (0.1,0.23)$) -- ($(p33) + (0,0.15)$) -- cycle;
  \coordinate (p34) at ($(C) + (1.33,2.23)$);
  \filldraw[fill=black] (p34) -- ($(p34) + (0.1,0.1)$) -- ($(p34) + (0.1,0.23)$) -- ($(p34) + (0,0.15)$) -- cycle;
  \coordinate (p35) at ($(C) + (1.33,1.91)$);
  \filldraw[fill=black] (p35) -- ($(p35) + (0.1,0.1)$) -- ($(p35) + (0.1,0.23)$) -- ($(p35) + (0,0.15)$) -- cycle;
  \coordinate (p36) at ($(C) + (1.13,2.87)$);
  \filldraw[fill=black] (p36) -- ($(p36) + (0.1,0.1)$) -- ($(p36) + (0.1,0.23)$) -- ($(p36) + (0,0.15)$) -- cycle;
  \coordinate (p37) at ($(C) + (1.13,1.91)$);
  \filldraw[fill=black] (p37) -- ($(p37) + (0.1,0.1)$) -- ($(p37) + (0.1,0.23)$) -- ($(p37) + (0,0.15)$) -- cycle;
  \node at ($(C) + (0.6,3.3)$) {Mask};
  \draw (D) to[bend right] ($(D) + (0,1.75)$) node[above] {Lens} to[bend right] (D);
	\draw[thick] ($(E) + (0.11,0.12)$) -- ($(E) + (1.5,1.2)$) -- ($(E) + (1.5,3.29)$) -- ($(E) + (0.11,2.2)$) -- cycle;
	\draw ($(E) + (0.3,0.28)$) -- ($(E) + (0.3,2.35)$);
	\draw ($(E) + (0.5,0.42)$) -- ($(E) + (0.5,2.5)$);
	\draw ($(E) + (0.7,0.58)$) -- ($(E) + (0.7,2.65)$);
	\draw ($(E) + (0.9,0.75)$) -- ($(E) + (0.9,2.83)$);
	\draw ($(E) + (1.1,0.9)$) -- ($(E) + (1.1,2.99)$);
	\draw ($(E) + (1.3,1.05)$) -- ($(E) + (1.3,3.15)$);
	\draw ($(E) + (0.11,0.4)$) -- ($(E) + (1.5,1.5)$);
	\draw ($(E) + (0.11,0.7)$) -- ($(E) + (1.5,1.8)$);
	\draw ($(E) + (0.11,1.0)$) -- ($(E) + (1.5,2.1)$);
	\draw ($(E) + (0.11,1.3)$) -- ($(E) + (1.5,2.4)$);
	\draw ($(E) + (0.11,1.6)$) -- ($(E) + (1.5,2.7)$);
	\draw ($(E) + (0.11,1.9)$) -- ($(E) + (1.5,3.0)$);
	\node at ($(E) + (0.75,3.4)$) {Sensor};
  }
\caption{Optical system for random mask imaging. Light from the scene is focused onto an occlusive programmable mask or digital micromirror device (DMD), which pointwise-modulates an image of the scene according to a programmed pattern. The modulated scene image is then focused (or blurred) onto a sensor. An image of the scene may be computationally reconstructed from a series of sensor measurements acquired using different mask patterns.}
\label{fig:setup}
\end{figure}
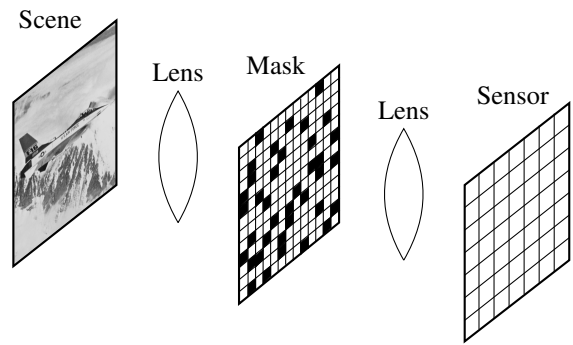

\begin{figure*}
\centering
  \tikz{
  		\coordinate (A) at (0,0);
  		\coordinate (B) at (2.1,0); 
  		\coordinate (C) at (5,0);   
  		\coordinate (D) at (6.5,0); 
  		\coordinate (E) at (9.4,0);  
  		\coordinate (F) at (11.7,0);  
  		\node at (A) [left] {\includegraphics[scale=0.075]{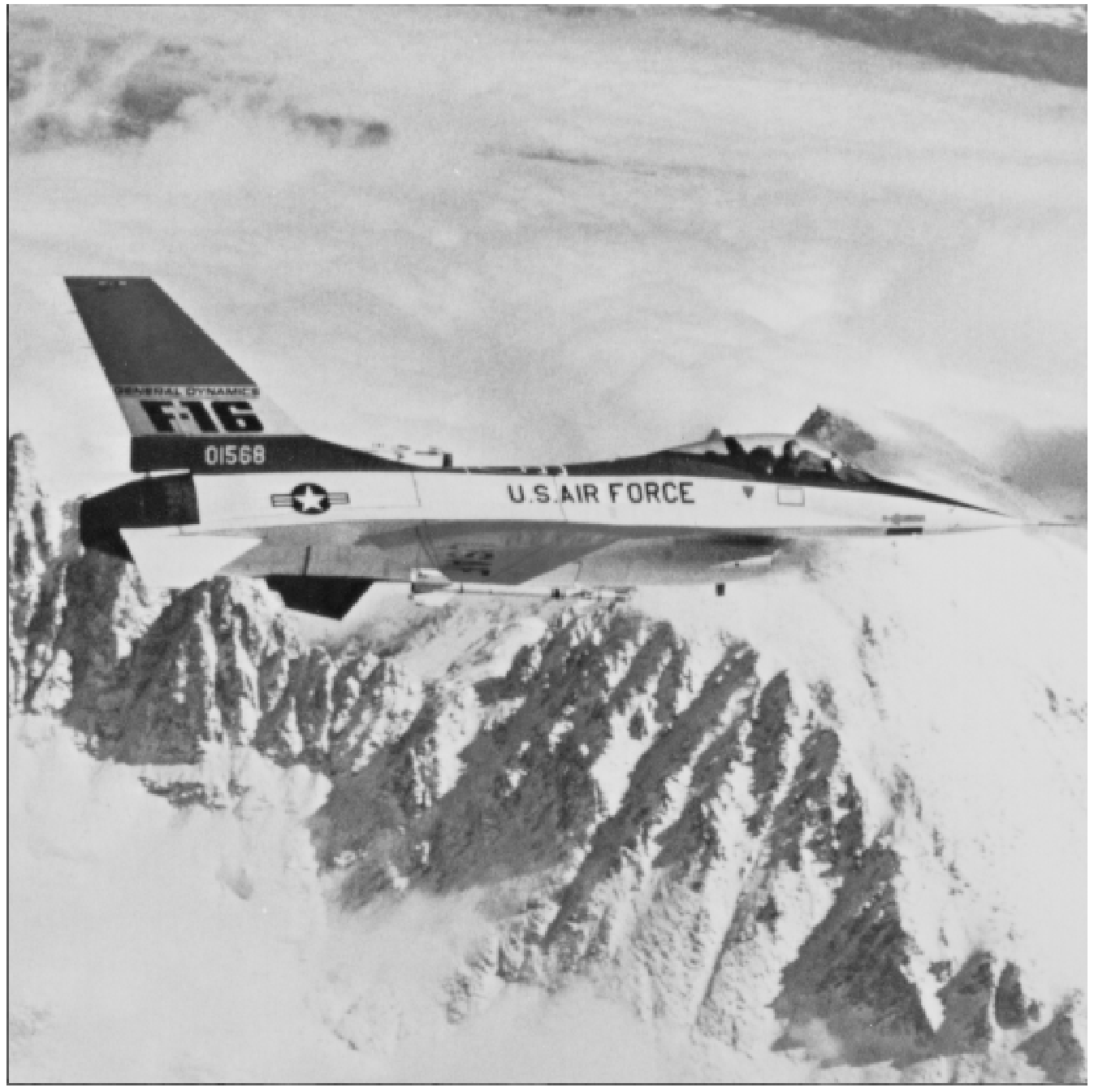}};
		\draw[very thick] [->] (A) -- node [midway,above=0.5em] {\large modulation} node [midway,below=0.5em] {\Large $\mathbf{D}_k$} (B); 
		\node at (B) [right] {\includegraphics[scale=0.075]{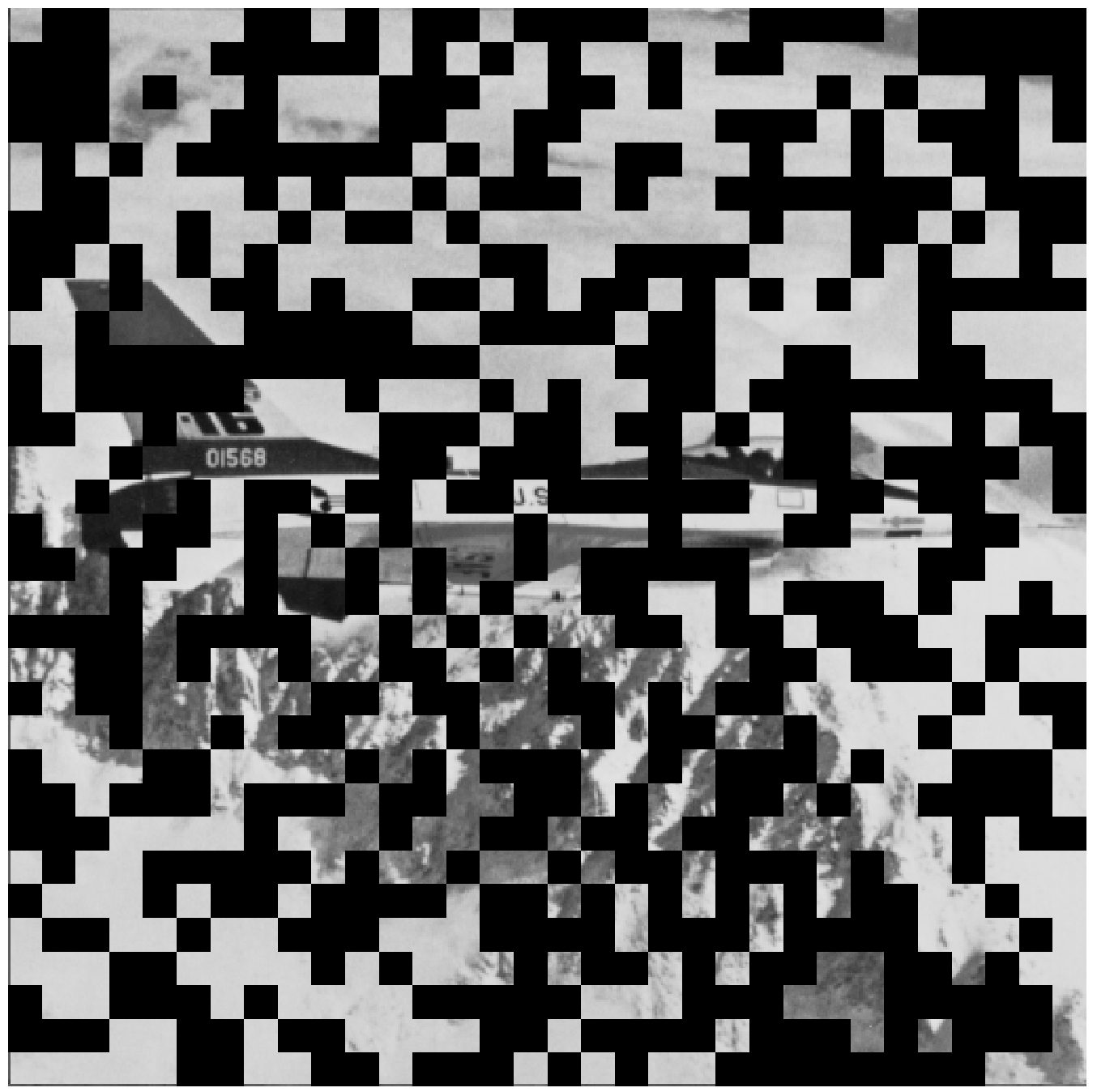}};
		\draw[very thick] [->] (C) -- node [midway,above=0.5em] {\large blurring} node [midway,below=0.5em] {\Large $\mathbf{B}_k$} (D);
		\node at (D) [right] {\includegraphics[scale=0.075]{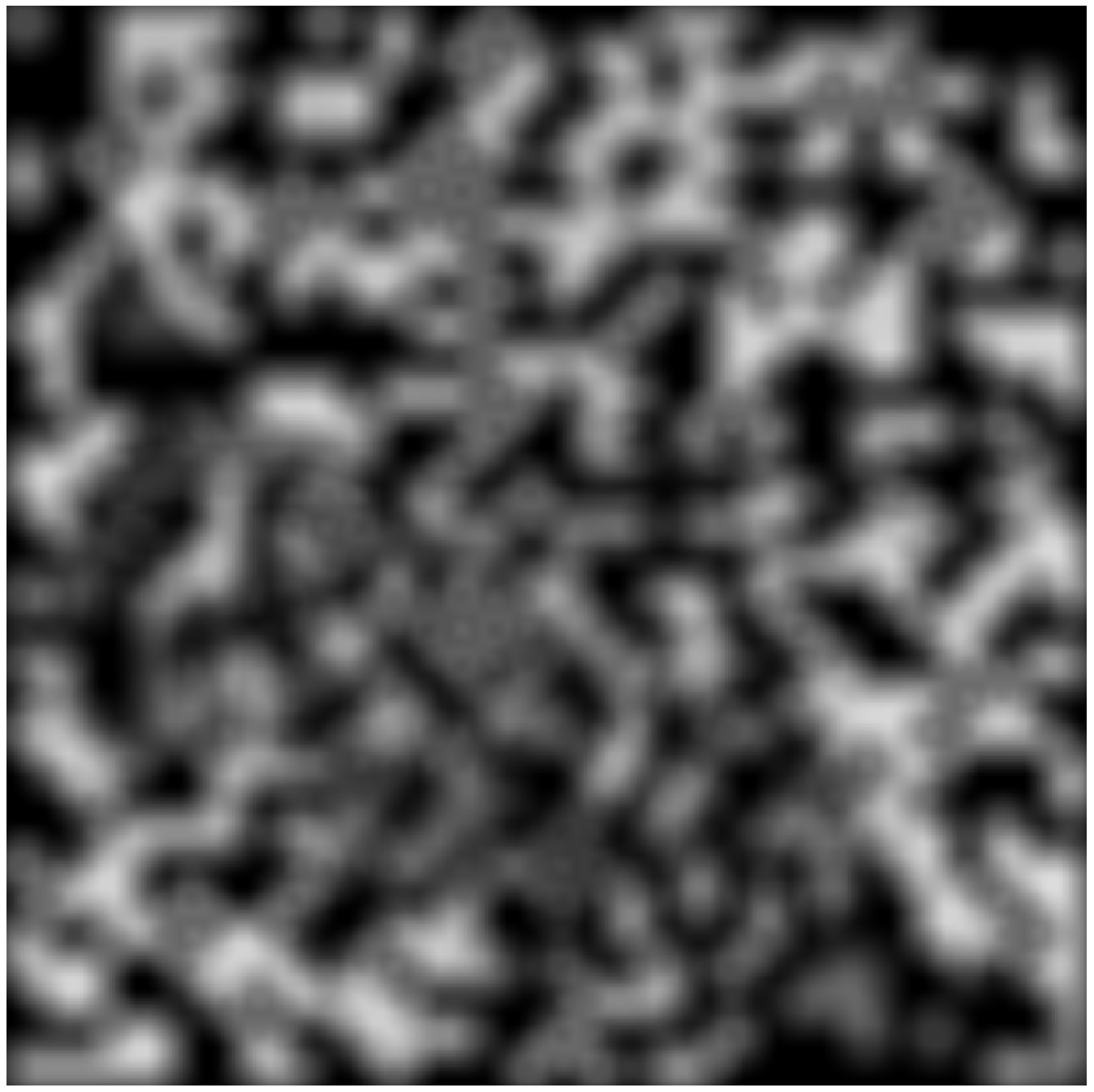}};
		\draw[very thick] [->] (E) -- node [midway,above=0.5em] {\large subsampling} node [midway,below=0.5em] {\Large $\mathbf{S}$} (F);
		\node at (F) [right] {\includegraphics[scale=0.16]{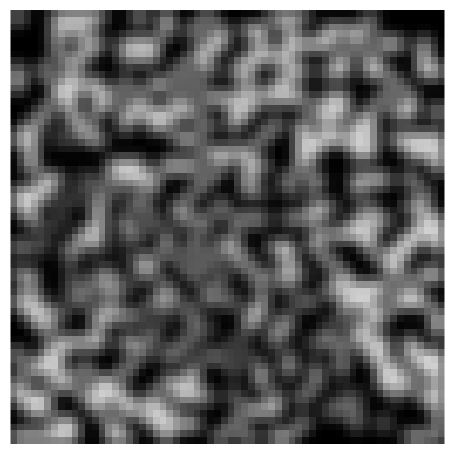}};
		\node at (-1.45, 1.6) {\Large scene};              
		\node at (-1.45,-1.7) {\Large $\mathbf{x}$};  
		\node at (3.6, -1.7) {\Large $\mathbf{D}_k \mathbf{x}$}; 
		\node at (8, -1.7) {\Large $\mathbf{B}_k\mathbf{D}_k \mathbf{x}$}; 
		\node at (12.75, 1.3) {\large measurement};
		\node at (12.9, -1.7) {\Large $\mathbf{y}_k$}; 
	}
\caption{How an example image is modified as it passes through a random mask imaging system. A scene image $x$ is pointwise modulated by a pattern matrix $D_k$ which represents the action of an occlusive programmable mask. The modulated image is then blurred, as represented by the operator $B_k$. This blurry, modulated image is then subsampled by the sensor, represented by the operator $S$, to obtain the measurement $y_k$.}
\label{fig:flow}
\end{figure*}
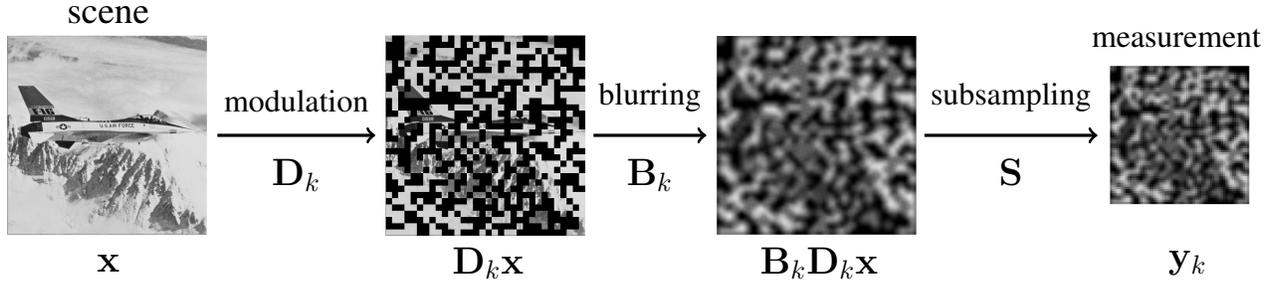

An illustration of a general random mask imaging system is shown in Figure \ref{fig:setup}. The basic setup contains two lenses, a programmable mask, and a sensor. The first lens focuses light from the scene onto the plane of an occlusive programmable mask, forming an image which the mask pointwise attenuates in blocks. A second lens blurs this modulated image onto the plane of the sensor, which integrates the light falling on each individual sensor element to produce a measurement. An image of the full scene may be computationally reconstructed from a series of such measurements taken with varying mask patterns. An example of how an image is modified as it moves through such a system is shown in Figure \ref{fig:flow}.

\subsection{Model}  \label{mathmodel}
Let us denote the resolution of the mask as $N$, the resolution of the sensor array as $M$, and the resolution of the image we aim to measure as $R$. We will use the vector $x \in \mathbb{R}^R$ to represent the $R$-pixel discrete approximation of the continuous two-dimensional image projected onto the mask by the first lens. Note that the scene may be three-dimensional, but ultimately we are trying to obtain a digital image that is a piecewise-constant approximation of whatever two-dimensional image is present on the plane of the mask. This image is assumed to be unchanging over the course of the measurement period. Here we are also implicitly assuming that the limiting factor in the resolution of a conventional imaging system using this mask and/or sensor array would be the resolutions of those components and not the optical resolving power of the first lens. 

A single measurement by the imaging system may be expressed as
\begin{equation} 
	\label{eq:measurement}
	y_k = \mathbf{S} \mathbf{B}_k \mathbf{D}_k x + \mathrm{noise},
\end{equation}
where $y_k \in \mathbb{R}^{M}$ is the vectorized output of the sensor array, $\mathbf{D}_k$ represents the operation of the mask, $\mathbf{B}_k$ represents the blur between the mask and the sensor, and $\mathbf{S}$ represents the sampling performed by the sensor.

Let us describe each of the matrices in this expression in more detail:
\begin{itemize}

\item $\mathbf{D}_k$ is an $R \times R$ binary diagonal matrix which captures how the mask point-wise modulates the pixels of the discrete approximation of the scene image $x$ on or off. If we let $d_k = \textbf{diag}(\mathbf{D}_k)$, then another way of expressing the action of the matrix $\mathbf{D}_k$ is $\mathbf{D}_k x = d_k \circ x$, where $\circ$ denotes entry-wise vector multiplication (the Hadamard product). Because there is a resolution mismatch between the image and the mask, the vector $d_k$ must be an $N$-pixel pattern scaled up to resolution $R$. In other words, the action of $\mathbf{D}_k$ is piecewise-constant over $c \times c$ blocks of pixels in the $R$-pixel image $x$, where $c = \sqrt{R/N}$.

\item $\mathbf{B}_k$ is an $R \times R$ matrix which represents the blurring operation of the second lens. If the system is spatially invariant, the rows of this matrix consist of different shifts of the vectorized point spread function.

\item $\mathbf{S}$ is an $M \times R$ sub-sampling matrix which sums together blocks of the blurry, modulated image to produce the measurement $y_k$. 
\end{itemize}

Note that the blurring matrices $B_k$ are indexed individually for notational convenience, although this is not meant to suggest the use of a different blur kernel for every measurement. In the presented results there are only a total of $1$ to $3$ unique $\mathbf{B}_k$ in each experiment. 

The sensor sampling matrix $\mathbf{S}$ can safely be assumed constant, so we combine it with $\mathbf{B}_k$ to form a general weighting matrix $\mathbf{W}_k = \mathbf{S} \mathbf{B}_k$ which describes how the image that passes through the programmable mask is converted to a measurement. We stack the set of measurement vectors $\{y_k\}_{k=1}^K$ column-wise to yield the vector $\mathbf{Y} \in \mathbb{R}^{KM}$ representing a complete set of $K$ measurements. The full linear model of the imaging system may then be written as 
\begin{equation} 
\label{eq:model}
\mathbf{Y} = 
\begin{bmatrix}
y_1\\ 
\vdots\\ 
y_K
\end{bmatrix}
=
\begin{bmatrix}
\mathbf{W}_1 &  &  \\ 
 & \ddots &  \\ 
 &   & \mathbf{W}_K
\end{bmatrix}
\begin{bmatrix}
\mathbf{D}_1\\ 
\vdots \\ 
\mathbf{D}_K
\end{bmatrix}
\begin{bmatrix}
 \\ 
x\\ 
\\
\end{bmatrix} 
=
\mathbf{A} x
\end{equation}

\subsection{Recovery using Least-Squares}
Given the system model $\mathbf{A}$ and the measurements $\mathbf{Y}$, we may estimate the discrete approximation of the scene image $x$ by solving the least-squares problem
\[
	\minimize_x~\frac{1}{2}\|\mathbf{A}x - \mathbf{Y}\|_2^2 + \delta\|x\|_2^2,
\]
the Tikhonov-regularized solution of which is
\begin{equation} 
	\label{eq:recovery}
	\hat{x} = (\mathbf{A}^\mathsf{T} \mathbf{A} + \delta\mathbf{I})^{-1} \mathbf{A}^\mathsf{T} \mathbf{Y}.
\end{equation}
where the regularization parameter $\delta>0$ is user-defined.  The effectiveness\footnote{Of course, another factor in the effectiveness of \eqref{eq:recovery} is how accurately  $\mathbf{A}$ models the optical system.  In practice, the accuracy of the reconstruction could vary depending on how we discretize the image.  But, as the experiments in Section~\ref{sec:opticsresults} support, using a straightforward pixelization is often sufficient.} of \eqref{eq:recovery} in estimating $x$ depends critically on the eigenvalue spectrum of the system matrix $\mathbf{A}^\mathsf{T} \mathbf{A}$ --- if there are $R$ nonzero eigenvalues bounded safely away from $0$, then we can confidently reconstruct the image for even small values of $\delta$.  Thus, we want to choose the matrices  $\{\mathbf{D}_k\}$ and $\{\mathbf{B}_k\}$ in such a way that $\mathbf{A}^\mathsf{T} \mathbf{A}$ is well-conditioned, where
\begin{equation} 
	\label{eq:AtA}
	\mathbf{A}^\mathsf{T} \mathbf{A} 
	= 
	\sum_{k=1}^K \mathbf{D}_k \mathbf{W}_k^\mathsf{T} \mathbf{W}_k \mathbf{D}_k.
\end{equation}

Both the matrix  $\mathbf{A}^\mathsf{T} \mathbf{A}$ and the vector
\begin{equation}
	\label{eq:AtY}
	\mathbf{A}^\mathsf{T} \mathbf{Y}
	=
	\sum_{k=1}^K \mathbf{D}_k \mathbf{W}_k^\mathsf{T} y_k,
\end{equation}
can be computed efficiently. Considered as a linear operator, $\mathbf{A}$ consists of modulations (point-wise multiplications), convolutions, and local averaging, all of which scale to very high resolutions.  Moreover, if the blurring matrices $\mathbf{B}_k$ use highly localized kernels (as they do in our numerical simulations below), then both $\mathbf{A}$ and $\mathbf{A}^\mathsf{T} \mathbf{A}$ will be sparse and can be explicitly constructed, held in memory, and applied directly even at large scale.

Our numerical simulations below take $32 \times 32$ measurements to $64 \times 64$ reconstructed images --- in this case, all of the matrix calculations (including the inversion) can be done directly, and the spectrum can be explicitly computed.  For larger-scale problems, \eqref{eq:recovery} can be computed using an iterative method (e.g.\ conjugate gradients), with the component parts of $\mathbf{A}$ and $\mathbf{A}^\mathsf{T}$ carefully implemented.

There is no doubt that more sophisticated reconstruction methods, particularly those those that take advantage of the expected structure in the scene (e.g. using total variation or sparsity as a regularizer), would result in better-quality reconstructions.  But what we are interested in here is how the placement of different optical components affect the matrix $\mathbf{A}^\mathsf{T} \mathbf{A}$; these are best manifested in the least-squares reconstruction \eqref{eq:recovery}, and so we use this throughout the paper.

\subsection{How Modulated Blurring Enables Superresolution}
\label{sec:blurringsuper}

\begin{figure}
	\centering
	\includegraphics[scale=0.35]{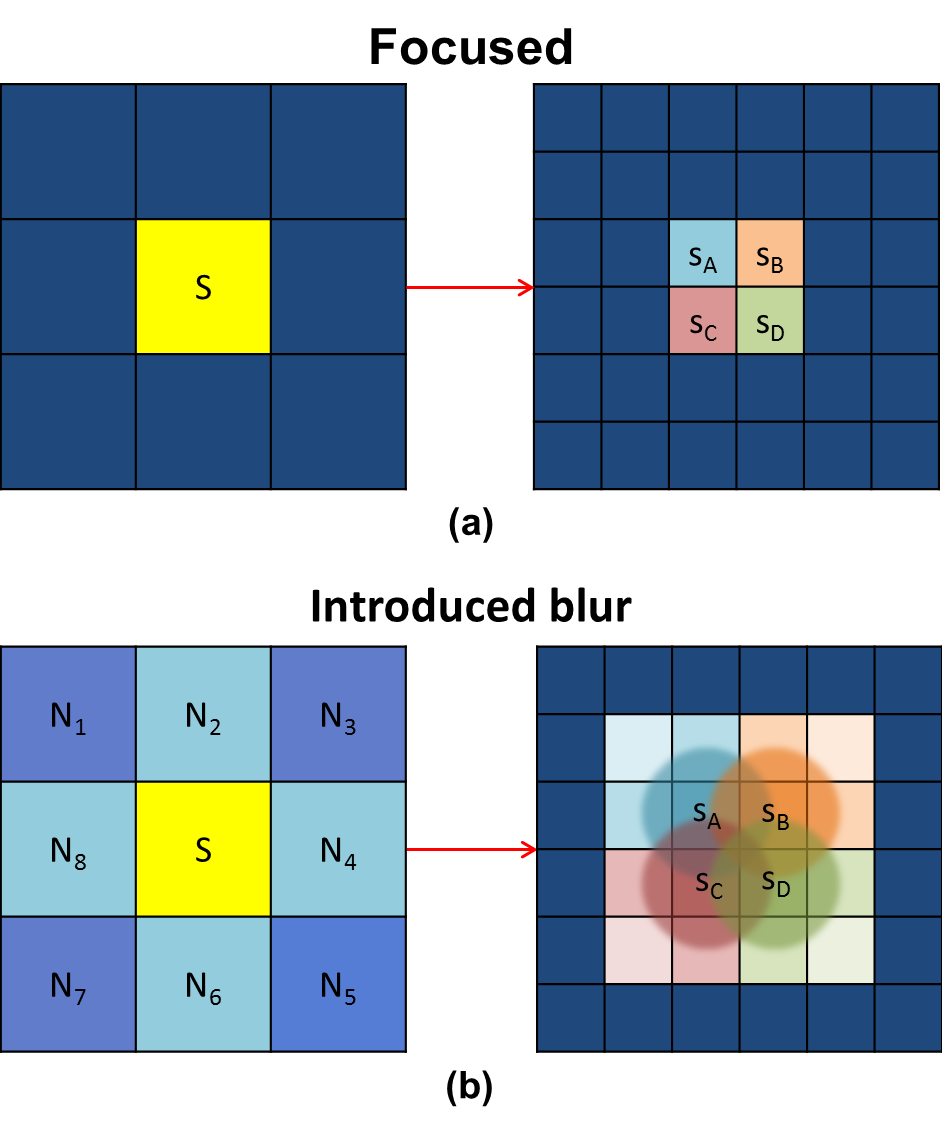}
	\caption{Top row: sensor response $S$ to different sub-pixels ($s_A, s_B, s_C, s_D$) when the system is in focus, Bottom row: sensor response to different sub-pixels when a blur is introduced between the mask and sensor. When a blur is introduced, the sensor $S$ and its neighboring sensors $N_i$ observe different responses to combinations of individual sub-pixels within one mask element.}
	\label{fig:HowBlurEnablesSuperres}
\end{figure}

To develop some intuition for how the introduction of a blur after the modulation gives us meaningful sub-pixel discernablility, we compare two cases as shown in Figure~\ref{fig:HowBlurEnablesSuperres}: (a) when the system is perfectly in-focus, and (b) when there is blurring between the mask and the sensor. 

Assume for both cases that the field of view of the sensor exactly covers the mask, and suppose we fix the resolution of the scene image $x$ to be $4N$, where $N$ is the resolution of the mask. Then if any single mask element is on while the rest are off, the modulated image contains exactly $4$ pixels. If we can determine the intensities of each of the $4$ pixels in each of the $N$ mirrors, then we will have effectively captured a $4N$ resolution image.

With the system in focus, all the rays passing through each open mask element strike a single sensor element. Regardless of what patterns or what number of patterns are displayed, each sensor element records only the total intensity of the light passing through one mask element, $S=s_A+s_B+s_C+s_D$, observing none of the sub-pixel variation. In this case, the resolution at which we can resolve is essentially the resolution of the mask. 

By introducing a blurring operation into the optical pathway, we are able to obtain more information about the sub-pixel regions through crosstalk of the sensors. Now multiple sensor elements can be made to measure different linear combinations of the sub-pixels, as shown in Figure \ref{fig:HowBlurEnablesSuperres}(b). If these linear combinations are diverse enough, we will be able to invert the resulting systems of linear equations to obtain the scene at a higher resolution. The switching of the DMD patterns over time is therefore effectively acting to disentangle portions of the scene that would normally be sensed concomitantly. By separating the contributions from neighboring pixels over a given set of measurements we are able to sense pixels of $x$ at a higher resolution than both the mask and sensor.

\subsection{An Example in 1D}

We can be even more explicit about how the combination of modulation and blurring helps us to superresolve by examining a one-dimensional example using a single blur kernel. Suppose $x$ is the ``high-resolution'' vector of length $R$ which we are trying to resolve using a mask of resolution $N = R/2$, and we use a blurring operator with impulse response $\begin{bmatrix} 1 & 2 & 1 \end{bmatrix}$. Then the matrices from section \ref{mathmodel} are 
\begin{align*}
\mathbf{S} &= 
\begin{bmatrix} 
1 & 1 & 0 & 0 & \cdots & 0 & 0 \\
0 & 0 & 1 & 1 & \cdots & 0 & 0 \\
\vdots &  &  &  &  &  & \vdots \\
0 & 0 & 0 & 0 & \cdots & 1 & 1 
\end{bmatrix}, \\
\mathbf{B} &= 
\begin{bmatrix}
2 & 1 & 0 & 0 & \cdots & 0 \\
1 & 2 & 1 & 0 & \cdots & 0 \\
\vdots & & \ddots & \ddots & \ddots & \vdots \\
0 & \cdots & & & 1 & 2
\end{bmatrix}, \\
\mathbf{D}_k &= 
\begin{bmatrix}
p_{1,k} & & & & & & \\
& p_{1,k} &&&&& \\
&& p_{2,k} &&&& \\
&&& p_{2,k} &&& \\
&&&& \ddots && \\
&&&&& p_{N,k} & \\
&&&&&& p_{N,k} \\
\end{bmatrix},
\end{align*}

where the values $p_{n,k}$ correspond to the pattern displayed on the mask during the $k^\text{th}$ measurement. To make the exposition simpler, at this point we will take $p_{n,k} \in \{-1,1\}$, and will note what changes when $p_{n,k} \in \{0,1\}$ at the end.

With the system in-focus, we have $\mathbf{W}_k = \mathbf{S}$ and the system matrix $\mathbf{A}^\mathsf{T} \mathbf{A}$ in \eqref{eq:AtA} is
\[
\sum_{k=1}^K \mathbf{D}_k \mathbf{S}^\mathsf{T} \mathbf{S} \mathbf{D}_k = 
\begin{bmatrix}
s_1 & s_1 & 0 & 0 & \cdots & & 0 \\
s_1 & s_1 & 0 & 0 & \cdots & & 0 \\
0 & 0 & s_2 & s_2 &  & & 0 \\
0 & 0 & s_2 & s_2 &  & & 0 \\
\vdots & \vdots &  &  & \ddots & & \vdots \\
 &  &  &  &  & s_N & s_N \\
0 & 0 & 0 & 0 & \cdots & s_N & s_N \\
\end{bmatrix}
\]
where
\[
s_n = \sum_{k=1}^K p_{n,k}^2
\]

Each of the $2 \times 2$ matrices along the block diagonal is at most rank one, so $\mathbf{A}^\mathsf{T} \mathbf{A}$ has rank at most $N$. Thus measurements with this system will contain no information about the image other than what is averaged over every mask element, and the least-squares solution in \eqref{eq:recovery} will be piecewise constant (as will the Tikhonov-regularized solution). 

\begin{figure*}
	\centering
	\begin{tabular}{ccc}
		\includegraphics[scale=0.33]{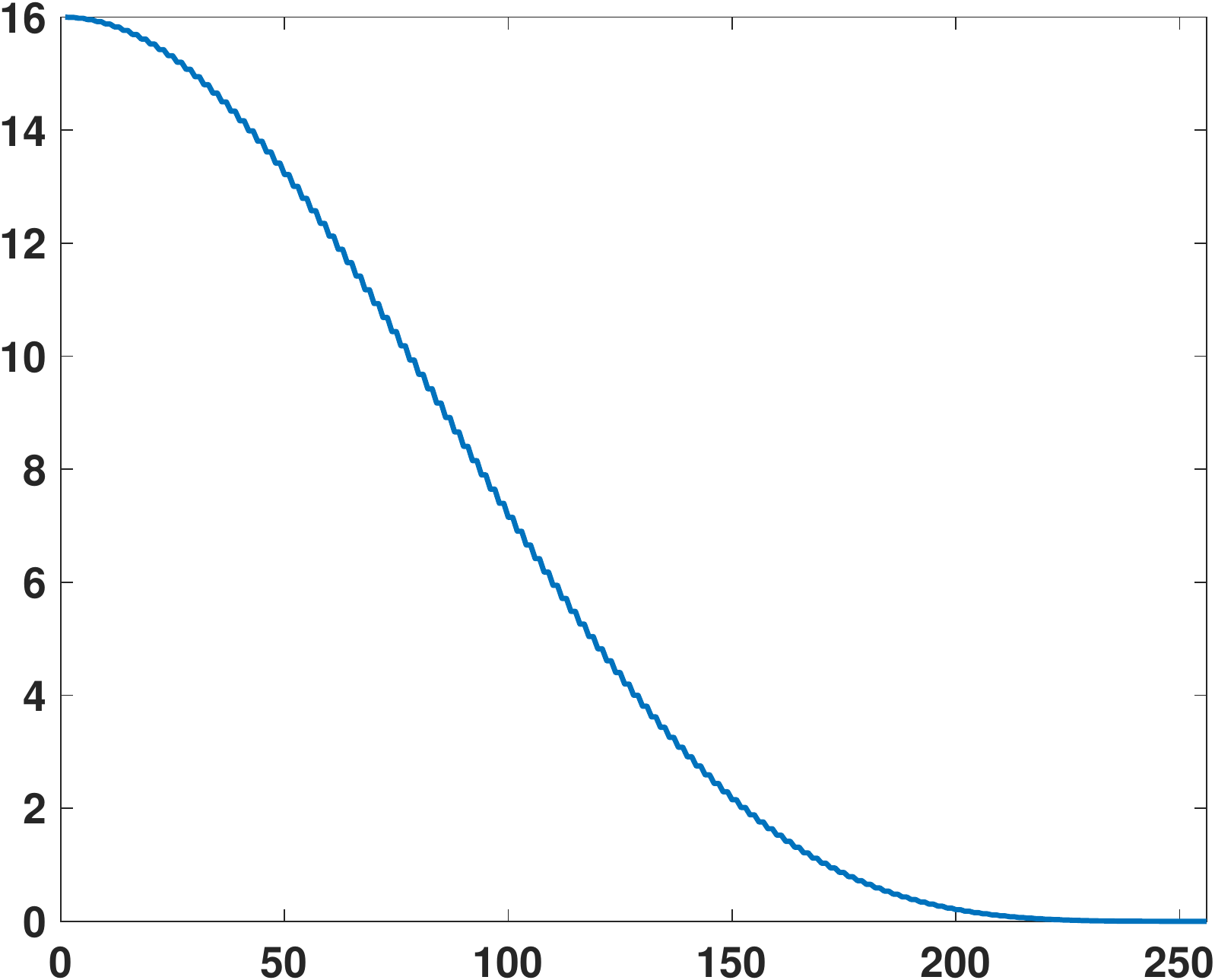} & 
		\includegraphics[scale=0.33]{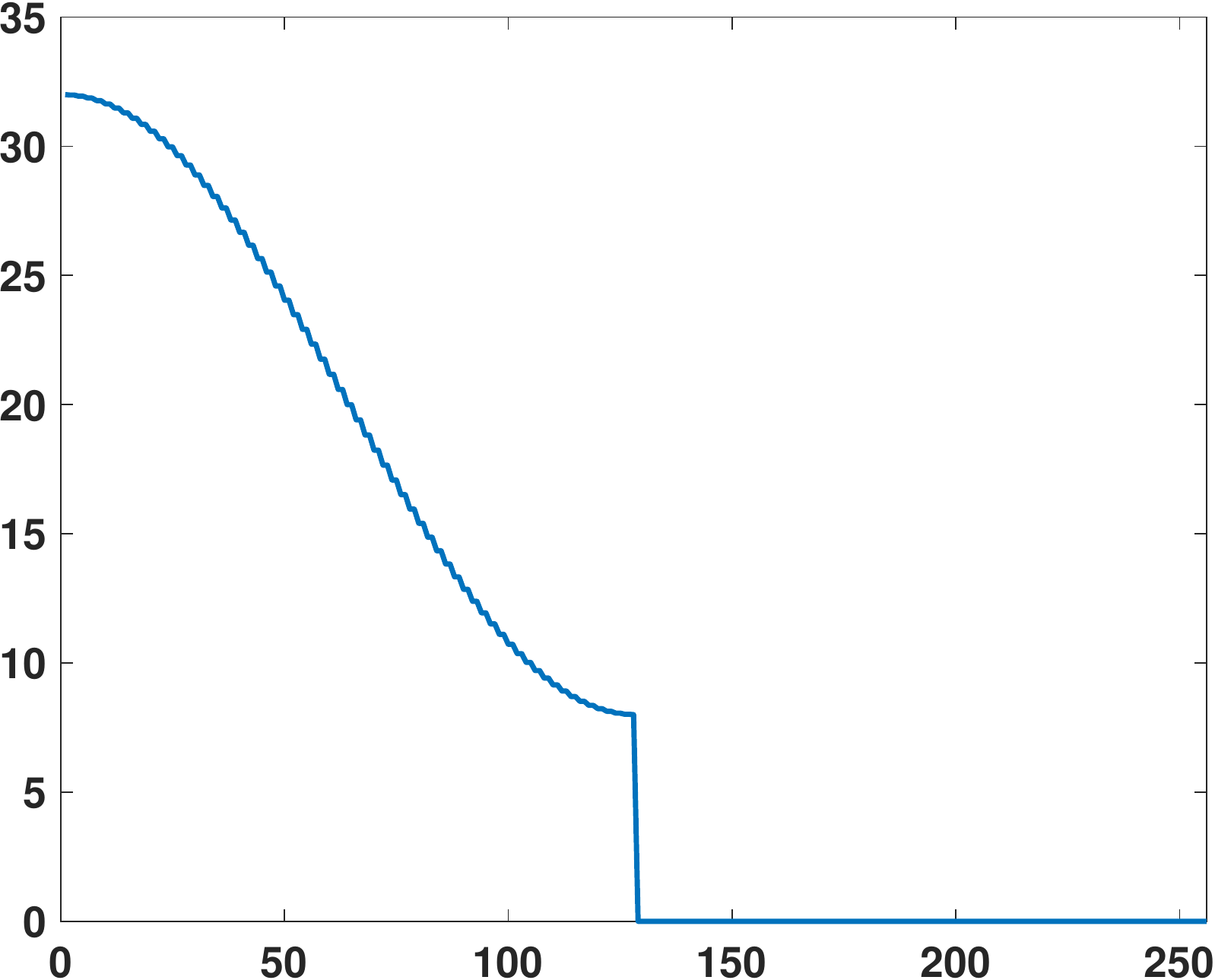} &
		\includegraphics[scale=0.33]{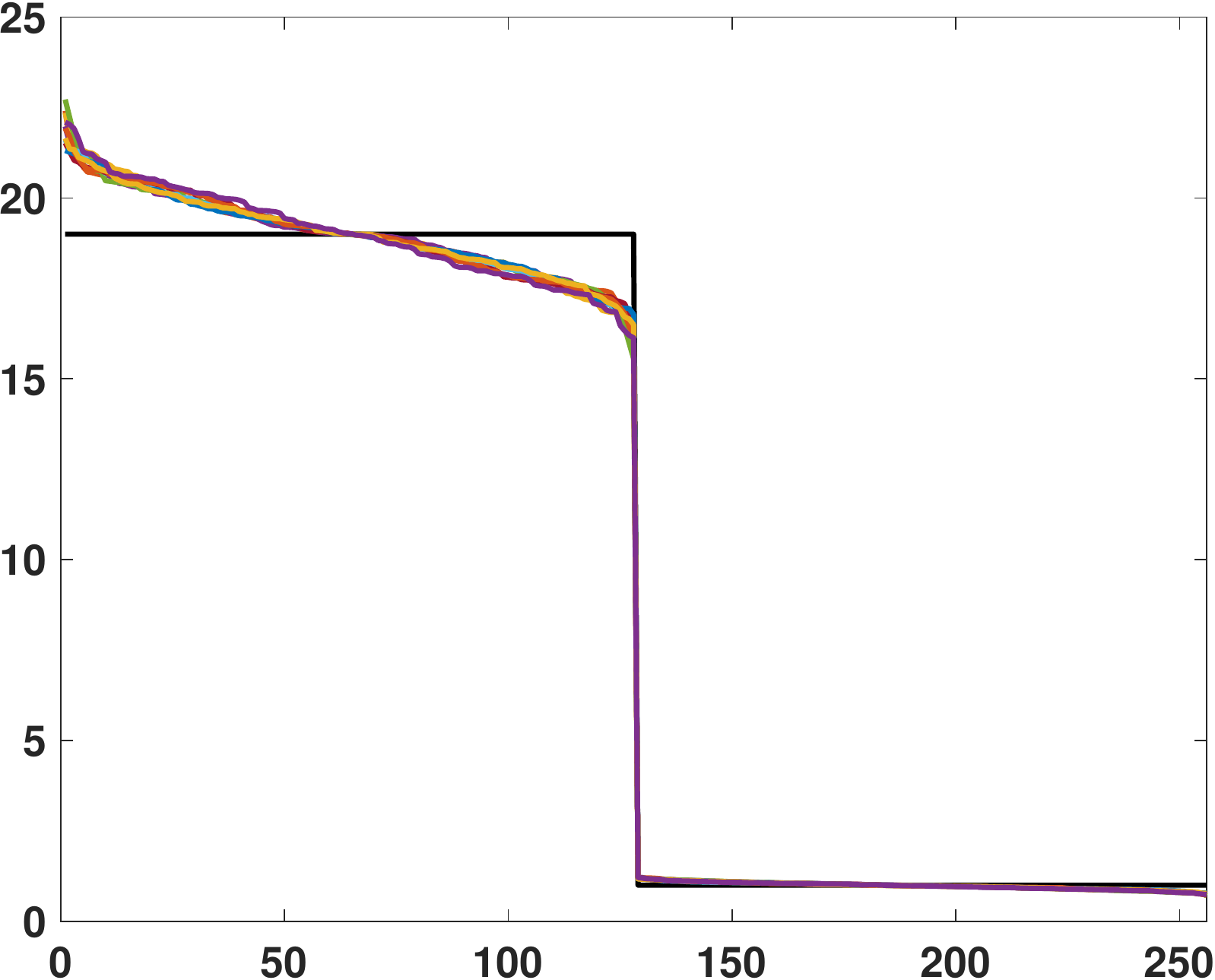} \\
		(a) & (b) & (c) \\
	\end{tabular}
	\caption{\small\sl The eigenvalue spectra for the key imaging matrices in the one-dimensional example.  Here $R=256$, $N=128$, and we are trying to super-resolve by a factor of two. (a) The eigenvalues of $\mathbf{B}^\mathsf{T}\mathbf{B}$.  This operator is essentially non-invertible over $\mathbb{R}^{256}$.  (b) The eigenvalues of $\mathbf{W}^\mathsf{T}\mathbf{W} = \mathbf{B}^\mathsf{T}\mathbf{S}^\mathsf{T}\mathbf{S}\mathbf{B}$.  This operator inherits the rank of $\mathbf{S}^\mathsf{T}\mathbf{S}$, and is only $128$. (c) The eigenvalues of $\E[\mathbf{A}^\mathsf{T}\mathbf{A}]$ are bi-level, shown by the black line.  The spectra for ten realizations of $\frac{1}{K}\mathbf{A}^\mathsf{T}\mathbf{A}$ for  $K=50$ are overlaid. All of these systems are full rank and stably invertible.}
	\label{fig:1Dspectrums}
\end{figure*} 

Initially it is not immediately obvious how adding a blur will improve the conditioning of this system, especially considering that $\mathbf{B}$ itself is not well-conditioned; its spectrum is shown in Figure~\ref{fig:1Dspectrums}(a). However, we will see that the interaction between the blurring and the mask patterns makes $\mathbf{A}^\mathsf{T} \mathbf{A}$ full rank. With the blurring included, the inner matrix $\mathbf{W}^\mathsf{T} \mathbf{W}$ in \eqref{eq:AtA} is
\[ 
\mathbf{B}^\mathsf{T} \mathbf{S}^\mathsf{T} \mathbf{S} \mathbf{B} = 
\begin{bmatrix}
9 & 9 & 3 & 0 & 0 & 0 & 0 & 0 & \cdots \\
9 & 10 & 6 & 3 & 1 & 0 & 0 & 0 & \cdots \\
3 & 6 & 10 & 9 & 3 & 0 & 0 & 0 & \cdots \\
0 & 3 & 9 & 10 & 6 & 3 & 1 & 0 & \cdots \\
0 & 1 & 3 & 6 & 10 & 9 & 3 & 0 & \cdots \\
\vdots &  &  &  &  & \ddots &  &  &  \\
\end{bmatrix}
\]
This matrix is at most rank $N = R/2$, since it cannot be higher rank than $\mathbf{S}$; its eigenvalue spectrum is shown in Figure~\ref{fig:1Dspectrums}(b).  However, when we include the modulation, the full system is given by \eqref{eq:AtA}, i.e. 
\[
\mathbf{A}^\mathsf{T} \mathbf{A} = \sum_{k=1}^K \mathbf{D}_k \mathbf{W}^\mathsf{T} \mathbf{W} \mathbf{D}_k
\]
The system matrix is now random, and for sufficiently many measurements $K$, we have the approximation $\mathbf{A}^\mathsf{T} \mathbf{A} \approx \E [ \mathbf{D}_k \mathbf{W}^\mathsf{T} \mathbf{W} \mathbf{D}_k ]$ (where we are omitting a constant factor of $K$). If we consider the expectation as approximated by a sum over increasingly many patterns, as $K$ increases the terms along the block diagonal remain unchanged, but those off the diagonal are divided by an increasing factor. Thus we obtain
\[
\E [ \mathbf{D}_k \mathbf{W}^\mathsf{T} \mathbf{W} \mathbf{D}_k ] = 
\begin{bmatrix} 
9 & 9 & & & & & & & \\
9 & 10 & & & & & & & \\
 &  & 10 & 9 & & & & & \\
 &  & 9 & 10 & & & & & \\
 &  &  & & 10 & 9 & & & \\
 &  &  & & 9 & 10 & & & \\
 &  &  & &  &  & \ddots & & \\ 
\end{bmatrix}.
\]
We see that the random coding, in the limit, makes the system matrix block diagonal and full rank. The eigenvalues of this matrix are simply the two eigenvalues of $\begin{bmatrix} 10 & 9 \\ 9 & 10 \end{bmatrix}$ (which are 19 and 1) repeated $N-2$ times, and the eigenvalues of $\begin{bmatrix} 9 & 9 \\ 9 & 10 \end{bmatrix}$ repeated twice on the boundaries. The spectrum of the expectation matrix is shown in Figure~\ref{fig:1Dspectrums}(c). The imaging system matrix has become full-rank, with a reasonable condition number of $\sim19$. 

Of course, for a finite number of measurements, the eigenvalues will not have exactly this simple two-value structure. But with a modest number of measurements, the spectrum of $\mathbf{A}^\mathsf{T} \mathbf{A}$ will have the same essential structure. In Figure~\ref{fig:1Dspectrums}(c) we plot 10 different realizations of $\mathbf{A}^\mathsf{T} \mathbf{A}$, each using $K=50$ patterns. The conditioning of each of these is similar to that of $\E [ \mathbf{A}^\mathsf{T} \mathbf{A}]$.


In practice, the entries in the $\mathbf{D}_k$ are better modeled as random variables with values in $\{0,1\}$. Of course, if a single measurement is taken with all of the $p_{n,\cdot}=1$, then other measurements can be subtracted digitally from this one, giving us $p_{n,k}\in\{-1,1\}$ as above.  If the system matrices are formed directly, then the expression for $\E [\mathbf{D}_k\mathbf{W}^\mathsf{T}\mathbf{W}\mathbf{D}_k]$ above is augmented by adding $1/4\mathbf{I} + 1/4\mathbf{1}\mathbf{1}^\mathsf{T}$, where $\mathbf{I}$ is the identity and $\mathbf{1}$ is the vector of all ones. As this matrix is positive definite, its addition will only increase the size of the smallest eigenvalue, and thus the eigenvalues of the expected system matrix are bounded away from zero.

Although we will not explore it further here, it may be possible to design the blurring kernel $\mathbf{B}$ to optimize the conditioning of the imaging system. In fact, a simple calculation shows that for a general symmetric length-3 filter $\begin{bmatrix} a & b & a \end{bmatrix}$ and the case of 2$\times$ superresolution, we have
\[
	\E[\mathbf{A}^\mathsf{T}\mathbf{A}] =
	\begin{bmatrix}
		a^2 + (a+b)^2 & (a+b)^2 & ~& \\
		 (a+b)^2 & a^2 + (a+b)^2 & ~ & \\
		 & & \ddots & ~
	\end{bmatrix}
\]
which has eigenvalues of $a^2+2(a+b)^2$ and $a^2$. The ratio of these takes the optimal value of one when $b=0$.

Stable super-resolution at larger factors is possible, but is more delicate.  As the super-resolution factor increases, the blurring kernel must get larger and more diverse.  Numerical experiments indicate that in this simple 1D scenario, the length of the filter must grow quadratically with the super-resolution factor.  The key $\sqrt{R/N} \times \sqrt{R/N}$ matrices involved also become more poorly conditioned, although they typically have $3$ or $4$ significant eigenvalues, indicating that stable super-resolution to these factors ($9\times$ or $16\times$) may be possible.

\section{Simulations}

With superresolution posed as a linear inverse problem, the superresolving capability of the system is ultimately dependent on the spectrum of the system matrix $\mathbf{A}^\mathsf{T} \mathbf{A}$. To validate the proposed method and better understand when we could expect to achieve good superresolution results, the system model was simulated under a range different conditions. In particular, we focused on identifying a blur kernel and a set of mask patterns that led to meaningful superresolution.

\subsection{Methodology}
The scene image formed at the plane of the mask was represented as an unchanging, discrete image of resolution $Q \times Q$, for $Q^2 \gg R$. For all simulations both the mask and sensor were fixed to be $32 \times 32$ arrays ($N = M = 1024$), and the aim was to superresolve by a factor of 4$\times$, i.e. to recover a $64 \times 64$ pixel image ($R = 4096$). The noise level was fixed at a level corresponding to a peak signal to noise ratio (PSNR) of $40$dB, which for the scene image used in the simulations corresponds to a noise standard deviation of $1.7$ pixel units.

The sensor was modeled as a square array of identical square elements flush with one another, and the magnification such that if the second lens were focused the masked image would map exactly into the sensor. Thus $\mathbf{S}$ was implemented by simple block-averaging. Assuming spatial invariance of the blur kernel, the $\mathbf{B}_k$ matrices were constructed as standard convolution matrices for each of the considered kernels. To ensure parity with respect to the mask, a single set of 500 32$\times$32 random binary patterns was used in all experiments. For $K < 500$, the same subset of this set of patterns was employed. The $\mathbf{D}_k$ matrices were formed by straightforwardly scaling the 32$\times$32 patterns up to resolution $Q \times Q$ and storing the vectorizations $d_k$ of these patterns (the diagonals of the $\mathbf{D}_k$). The product $\mathbf{D}_k x$ was then implemented as the pointwise product $d_k \circ x$. 

Given the matrix $\mathbf{S}$ and the matrix sets $\{\mathbf{B}_k\}_{k=1}^K$ and $\{\mathbf{D}_k\}_{k=1}^K$, a set of measurements $\{\mathbf{y}_k\}$ was simulated by applying the linear operators representing the actions of the mask, blur, and sensor in sequence to the scene image as in \eqref{eq:measurement}, then adding noise. 

Given the set of measurements, we first computed the matrix $\mathbf{A}^\mathsf{T} \mathbf{A}$ as in \eqref{eq:AtA} and vector $\mathbf{A}^\mathsf{T} \mathbf{Y}$ as in \eqref{eq:AtY}. The superresolved estimate was then obtained by directly solving \eqref{eq:recovery}, for some Tikhonov parameter. In the results that follow, we report error metrics obtained when using the optimal (in terms of the mean of the squared errors) Tikhonov factor. In practice this parameter may be easily tuned for a specific application.

\subsection{Choice of Blur Kernel}

\begin{figure}
	\centering
	\includegraphics[scale=0.25]{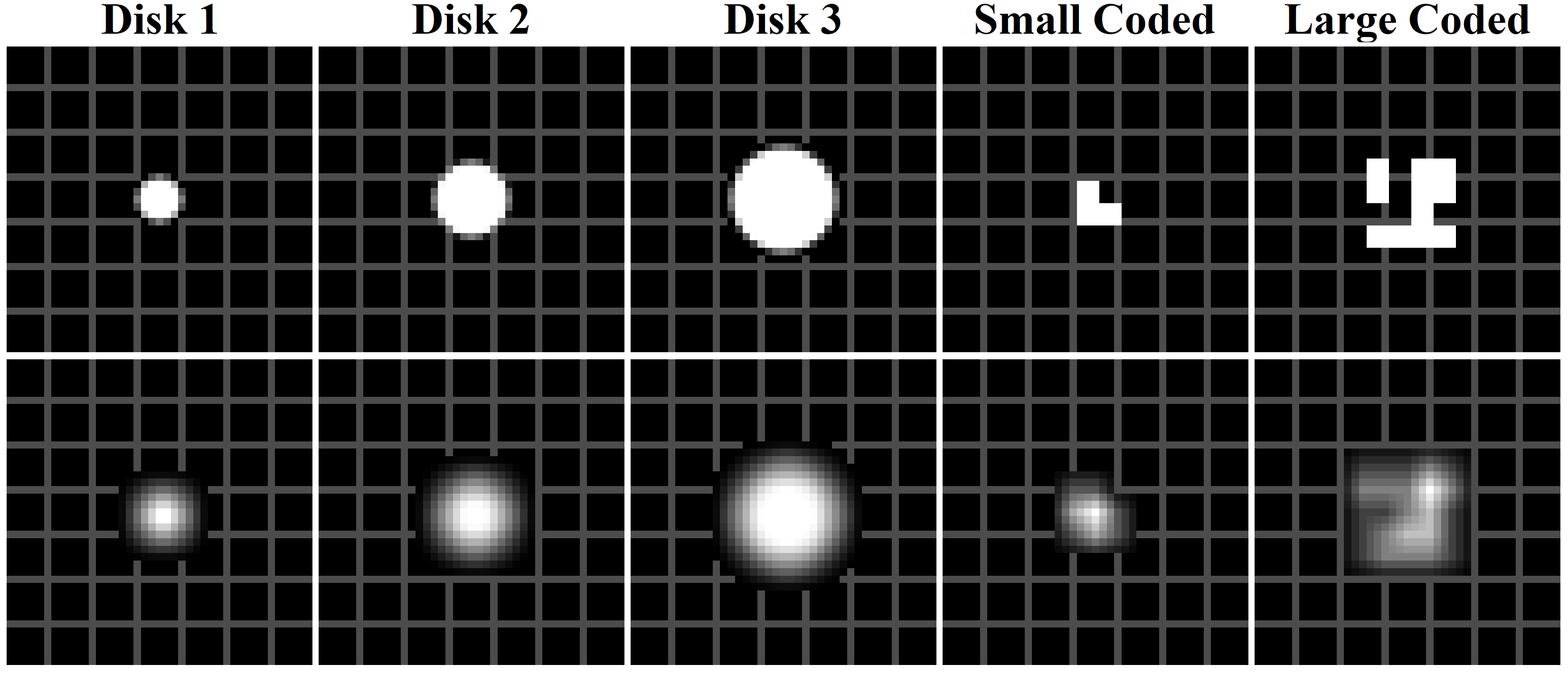}
	\caption{Example blur kernels considered in simulation, including disks of several sizes as well as coded kernels of different sizes, patterns, and resolutions. Grid lines indicating the divisions between sensor pixels are shown for scale. Below each kernel is the response observed at the sensor when exactly one mask element is open.}
	\label{fig:PSFSMR}
\end{figure}

Once the idea of introducing a blur between the mask and sensor is established, a natural next question is what particular blur kernel is best suited for superresolution. Towards identifying a reasonable kernel, the system was simulated using a number of different kernels, examples of which are shown in Figure \ref{fig:PSFSMR}.

\begin{figure}
	\centering
	\includegraphics[scale=0.27]{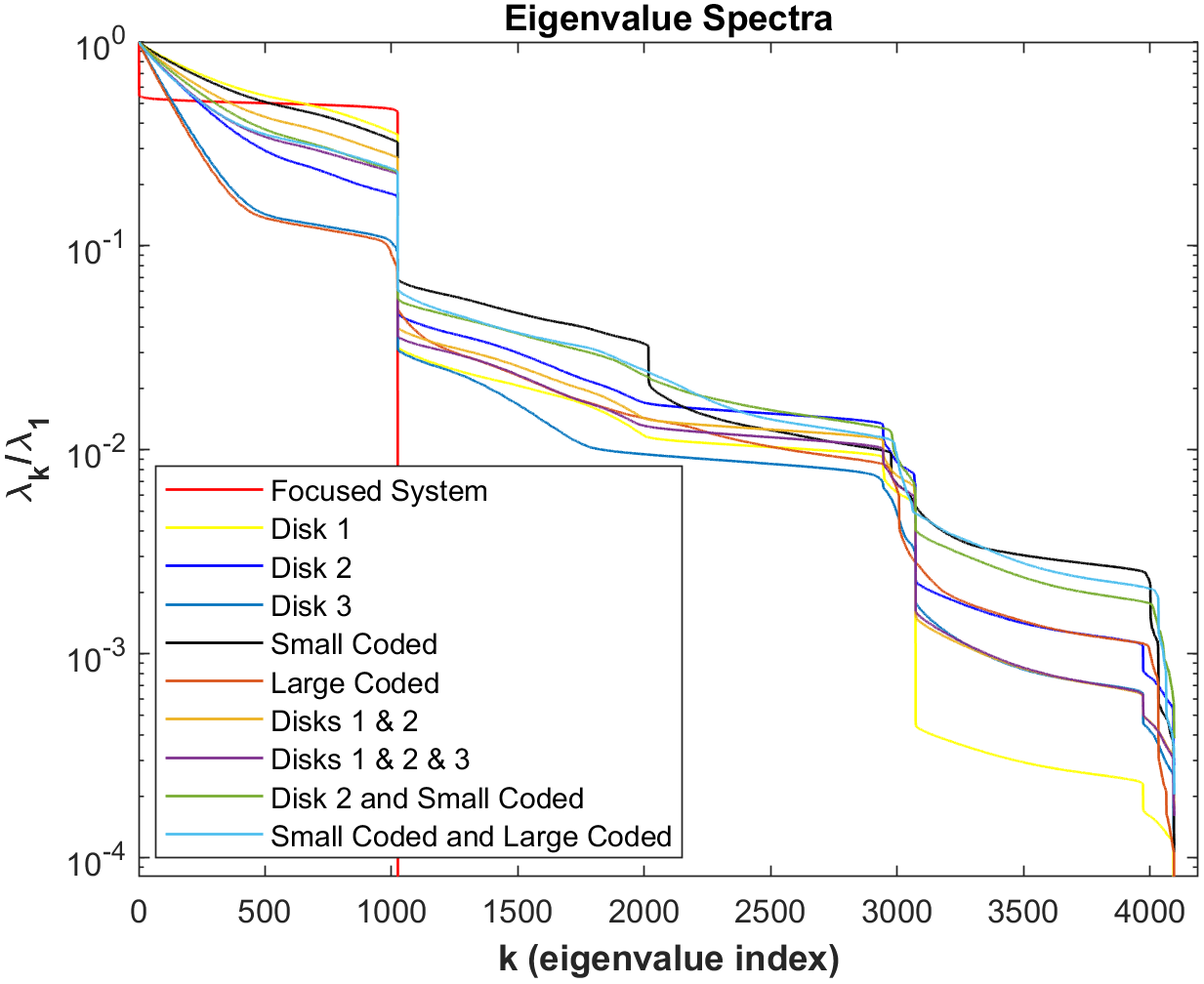}
	\caption{Eigenvalue spectra of the full system matrix $\mathbf{A}^\mathsf{T} \mathbf{A}$ when using different blur kernels (or combinations of multiple blur kernels) for the second lens, plotted with the y-axis in log-scale. Because what matters is not the absolute values of the eigenvalues $\lambda_k$, but their ratios with respect to the largest eigenvalue, every spectrum was divided by its largest eigenvalue (denoted $\lambda_1$). The variance of the noise will determine the number of components that can be robustly resolved (and hence the effective super-resolution factor). Note that to create a coded kernel some amount of light is necessarily blocked by the coding mask, meaning coded kernels will tend to perform worse in practice in comparison to a simple disk kernel for an equivalent number of measurements.}
	\label{fig:spec}
\end{figure}

As previously mentioned, the superresolving capability of the system (across all SNRs) is ultimately a function of the spectrum of the system matrix $\mathbf{A}^\mathsf{T} \mathbf{A}$. Plotted in Figure \ref{fig:spec} are the spectra of systems employing some representative kernels and combinations of multiple kernels that were studied. If we were attempting to resolve at the resolution of the sensor/mask, we would be interested only in the largest $M = 1024$ eigenvalues. In this case it is clear that the in-focus system performs best. However, given that we are trying to superresolve, the portion of the spectrum that interests us is what lies beyond the first $M$ eigenvalues. Here we see that for an in-focus system all eigenvalues after the first $M$ are zero, showing that the in-focus system has no superresolving capability. Different blur kernels lead to systems with different spectra, and a comparison of the different cases shows that the overall best-performing PSF among those considered was a disk with diameter slightly less than that of two sensor pixels. Coded kernels can lead to better conditioned systems than larger or smaller disk kernels, however using an occlusive mask to code the second lens necessarily results in a loss of light, and coded systems generally perform slightly worse than systems using a simple blur (for an equivalent number of measurements).

\subsection{Choice of Mask Patterns}
 
Another way of thinking about the superresolution problem presented here is as a set of $N$ sub-problems, one for each mask element, in each of which the aim is to estimate the sub-pixels that a particular mask element co-modulates. If a single mask element is open, the sensor measures only linear combinations of the sub-pixels which that mask element interacts with. 

\begin{figure*}
	\centering
	\includegraphics[scale=0.34]{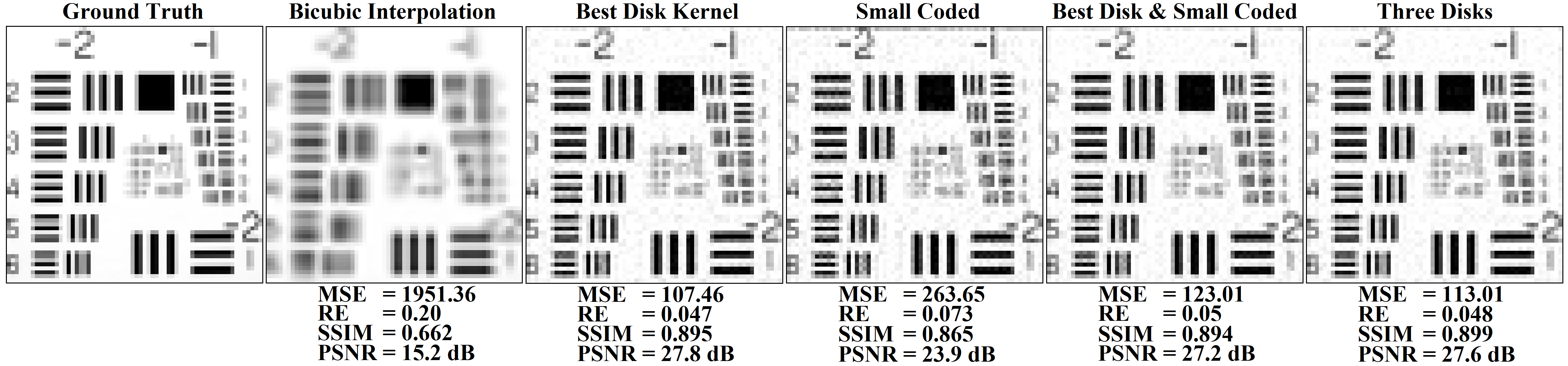}
	\caption{Example results simulated using different blur kernels, where we attempt to superresolve from a $32 \times 32$ image to a $64 \times 64$ image using $K = 100$ random mask patterns/measurements taken at a PSNR of $40$dB. The mean of the squared errors (MSE), relative error (RE), structural similarity index (SSIM), and peak signal-to-noise ration (PSNR) are computed for each result relative to the ground truth image. Also shown for comparison is a bicubic interpolation of the downsampled $32 \times 32$ image to $64 \times 64$.}
	\label{fig:SimResults}
\end{figure*}

When multiple mask elements are open simultaneously, each sensor element in general measures a linear combination of sub-pixels from multiple mask elements, and this mixing effect will in general make the recovery worse-conditioned. The "spectrum upper bound" for a particular kernel would be obtained when only one mask element is open in each measurement, and the spectrum degrades monotonically from this upper bound as we use patterns with more mask elements open in each measurement. Choosing a set of patterns is thus to some extent application-dependent, determined by temporal and noise constraints. In what follows we employ random mask patterns with half of the mask elements open in each measurement, which performs well in both simulations and practice.

\subsection{Simulation Results}

After simulating the measurements for a specified blur kernel and number of patterns, we recover the superresolved scene by Tikhonov-regularized least squares \eqref{eq:recovery}. For every case we report error metrics obtained after identifying and using the Tikhonov parameter that minimizes the mean of the squared errors (MSE) between the result and the ground truth. 

Example results for different kernels are shown in Figure \ref{fig:SimResults}. Systems employing a simple defocus performed better overall in terms of the error metrics relative to the coded kernels. Note that the edges of the images tend to be noisier and blurrier than their centers, because light from the edge pixels is measured at fewer total sensors. Because of the asymmetry of coded kernels particular edges can be worse-conditioned than others, such as the top edge in the first coded system shown in the figure. Truncating the edges results in better error metrics, but still not better than those obtained when using the best-performing disk kernel.

\begin{figure}
	\centering
	\includegraphics[scale=0.27]{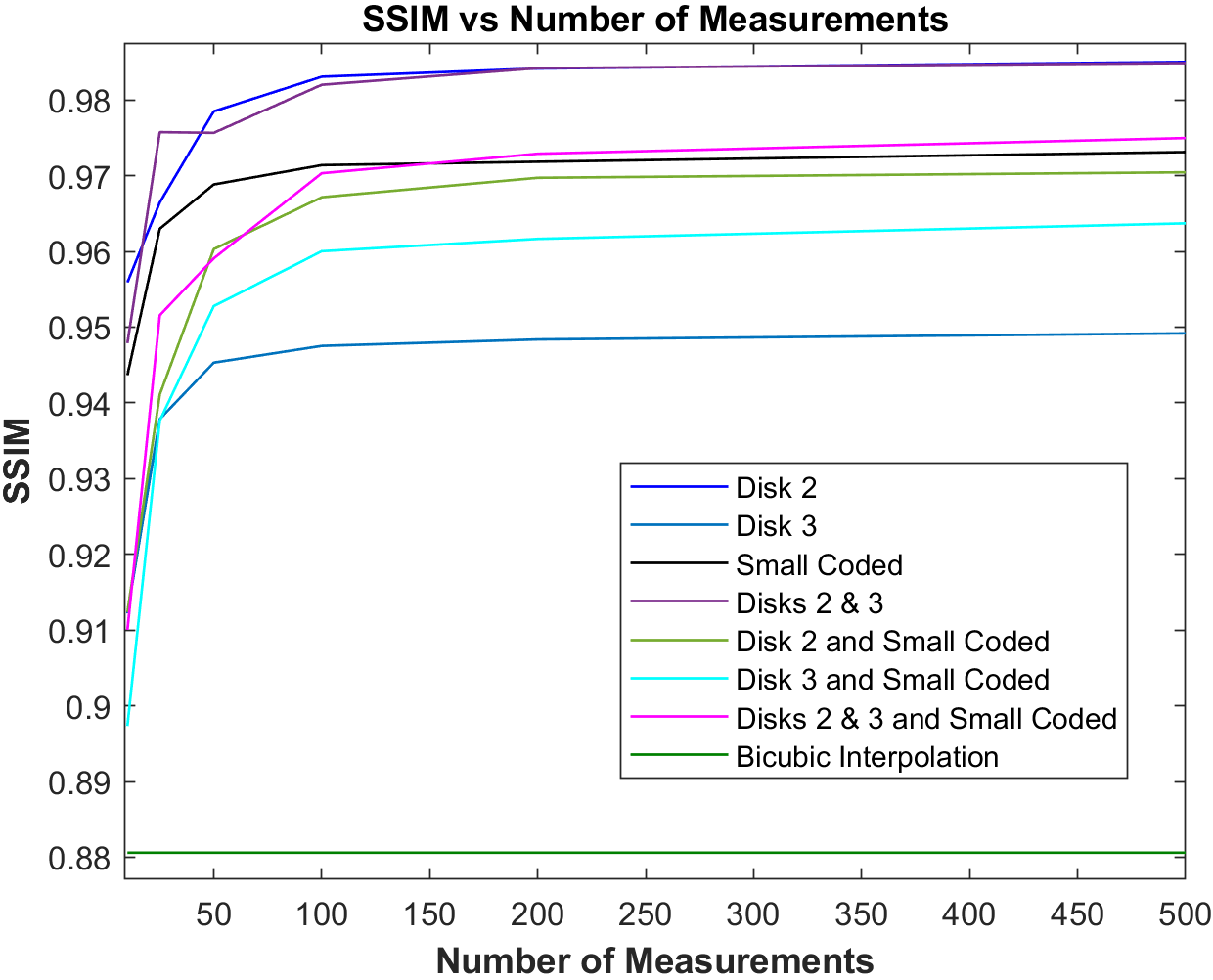}
	\caption{Simulated performance plots of structural similarity index (SSIM) of each system on an example image for different numbers of measurement. The bicubically interpolated image SSIM is given by the green line at the bottom. The red line along the bottom shows the SSIM for a random mask imaging system when the system is in-focus. The dark green line just above that represents the SSIM of the bicubically interpolated low-resolution image. The kernel that performed best in terms of SSIM was the disk-shaped kernel with diameter $1 \frac{2}{3}$ sensor-pixels.}
	\label{fig:SSIMvsKsimulated}
\end{figure}

Figure \ref{fig:SSIMvsKsimulated} shows the structural similarity (SSIM) \cite{SSIMeval} of the recovered image from systems using different kernels and a varying number of measurements in the case of a natural image. The performance improves rapidly for all kernels before leveling off around about 100 measurements. Notably all systems employing a blur demonstrated meaningful superresolution capability. Overall, the best performance was obtained with a blur kernel with a diameter of $1 \frac{2}{3}$ sensor pixels wide, although kernels between 1.5 and 2 sensor pixels in diameter performed similarly, so that is the range we targeted in the construction of the system.


\section{Real-world Implementation}

\subsection{Optical Layout}

\begin{figure}[ht]
	\centering
	\includegraphics[width = 8cm]{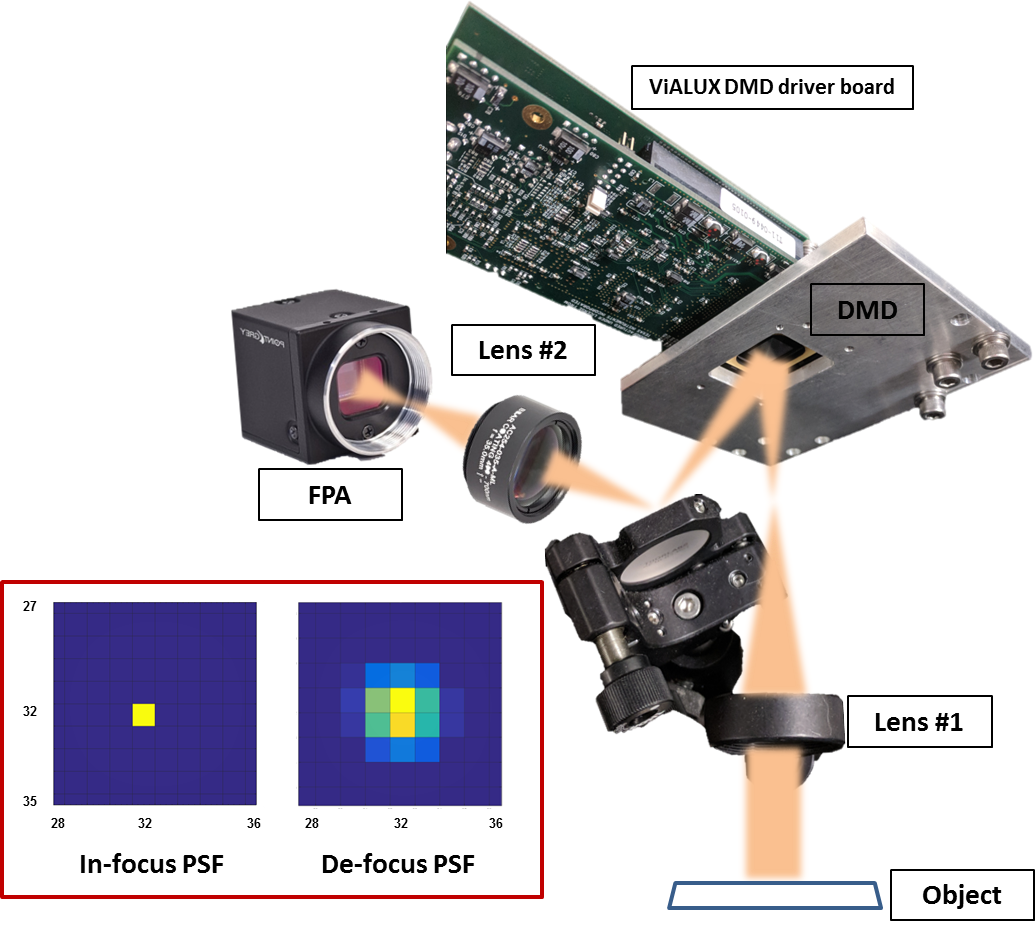}
	\caption{Illustration of the experimental setup. The sensor response when a single mirror of the DMD is in the ON state is shown for both the focused and blurred systems.}
	\label{fig:exp_setup}
\end{figure}

After validating the concept in simulations, we designed and built an optical system to achieve superresolved images in the laboratory. The experimental setup used to perform the superresolution procedure is illustrated in Figure \ref{fig:exp_setup}. A ViALUX DLPC410 digital micromirror device (DMD) chipset was used as the light modulator/image encoder to display the mask patterns. The DMD consists of hundreds of thousands of micromirrors whose orientations can be precisely controlled to tilt either left or right of the surface normal into ON or OFF states, either reflecting light towards the sensor or away from it, respectively. The ON/OFF values of the individual micromirrors of the DMD directly correspond to the elements of the binary pattern matrices $\{\mathbf{D}_k\}$. The highest switching speed of the DMD is in the tens of kilohertz, allowing for very fast optical modulation. In the implemented system, the switching speed was 100 Hz to coincide with the frame rate of the focal plane array, which was a Gazelle camera from Point Grey.

The pseudorandom binary patterns displayed on the DMD were of resolution $32 \times 32$, where at each measurement only half of the total DMD mirrors are turned to ON. Since the center portion of the DMD used was resolution 512$ \times $512, the target resolution was achieved by turning groups of 16$ \times $16 mirrors ON or OFF simultaneously. The scene was focused onto the DMD with a lens, then modulated by the pattern on the DMD, and then this modulated image was projected onto a CMOS focal plane array (FPA) through a second lens. The FPA was placed slightly behind the image plane of the second lens to implement the blurring operation. The distance between the image plane and the FPA determines the effective blur kernel size, and thereby the achievable superresolution performance. For the case of coded kernels, we simply put a coded mask in front of the second lens. The mask we implemented was the same as the small coded shape in Figure\ref{fig:PSFSMR}, a simple $2\times2$ binary pattern that has a size of $7\times7$ mm$^2$. The FPA employed had a native resolution of $2048 \times 1048$; in order to simulate the low-resolution sensor array, we combined blocks of sensor pixels in the center of the FPA together to obtain an effective resolution of $32 \times 32$. For this experiment we used the approximate blur kernel size discovered in simulations which produced the best results for 4$\times$ superresolution, so the reconstructed superresolved images have a resolution of $64 \times 64$.

\subsection{Calibration}

Ideally the blur kernel can be estimated based on the distance between the second lens and the FPA. However the imaging system also involves distortion and an imperfect mapping between the micromirrors of the DMD and the sensor pixels of the FPA. Thus the assumption of a spatially invariant blur was violated, and the purely mathematical estimation of the matrices $\mathbf{W}_k$ was not sufficient for a good reconstruction of the high-resolution image.

To obtain an estimate of the matrix $\mathbf{W}_k = \mathbf{S} \mathbf{B}_k$ in the forward model \eqref{eq:measurement}, we performed a point spread function (PSF) calibration process for higher density pixels on the DMD while maintaining the total number of micromirrors being used (512$ \times $512). This was achieved by turning on blocks of 8$ \times $8 pixels on the DMD to capture the response of our imaging system to point sources of effective resolution of 64$ \times $64. In the red box within Figure \ref{fig:exp_setup} are the PSFs of a perfectly focused system and a blurred system obtained through calibration. The $\mathbf{W}_k$ matrices acquired through the calibration process contain all the information about the optical system, including the blur kernel, optical distortion, and the precise mapping between the DMD and FPA. We can then use the estimated $\mathbf{W}_k$ to reconstruct the superresolved image from the randomly modulated measurements.

\subsection{Results}
\label{sec:opticsresults}

\begin{figure*}[ht]
	\centering
	\includegraphics[scale=0.34]{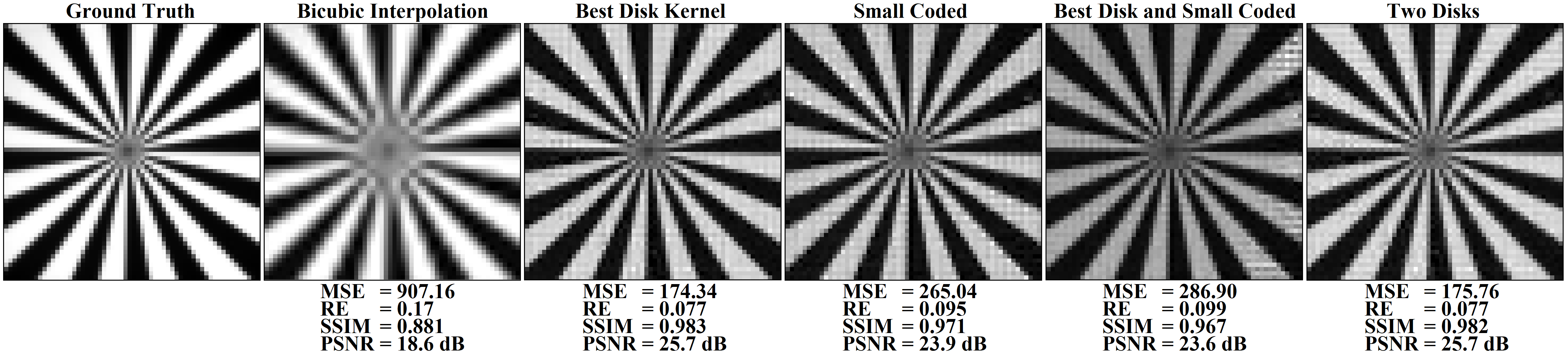}
	\caption{Example results computed from 100 measurements taken with the system using different blur kernels. The Tikhonov parameter yielding the optimal MSE was used for each kernel. Error metrics are the mean of the squared errors (MSE), relative error (RE), structural similarity index (SSIM), and peak signal-to-noise ratio (PSNR).}
	\label{fig:exp_res}
\end{figure*}

We directly solved the Tikhonov-regularized linear system to reconstruct the super-resolved image at resolution 64$\times$64 resolution from a series of random measurements taken with a 32$\times$32 resolution DMD and a 32$\times$32 resolution sensor array. Figure \ref{fig:exp_res} shows several superresolved images obtained for different blur kernels using 100 DMD patterns. Error metrics show that the proposed method meaningfully enhances the resolution of the imaging system beyond the resolution of both the mask and sensor array.

For every kernel, both a nonuniform illumination correction and a magnification and translation correction were performed for the sake of fair comparison. The nonuniform illumination correction was performed by first acquiring an image of a blank piece of paper with all the mirrors of the DMD set to ON, representing the effective background illumination pattern for the kernel. The obtained superresolved result was then divided by a version of this background pattern that was normalized to the $[0,1]$ range. For multiple kernels the background illumination pattern corresponded to the average of the individual kernel backgrounds. This correction compensates for the varying illumination observed in the scene that is created by the presence of the blur kernel. 

A magnification and translation correction was also performed after it was discovered that using different blur kernels introduced slightly different magnifications and translations, complicating fair comparison to a single ground truth image. To correct for these effects, for each kernel the ground truth image was simultaneously magnified and translated appropriately by trimming a small number of pixels from the edges of the 512$\times$512 target image before down-sampling to the 64$\times$64 resolution. For each kernel the cropping that yielded the minimal MSE is reported.

\section{Analysis}

\subsection{Performance Characterization}

\begin{figure}
	\centering
	\includegraphics[width = 9cm]{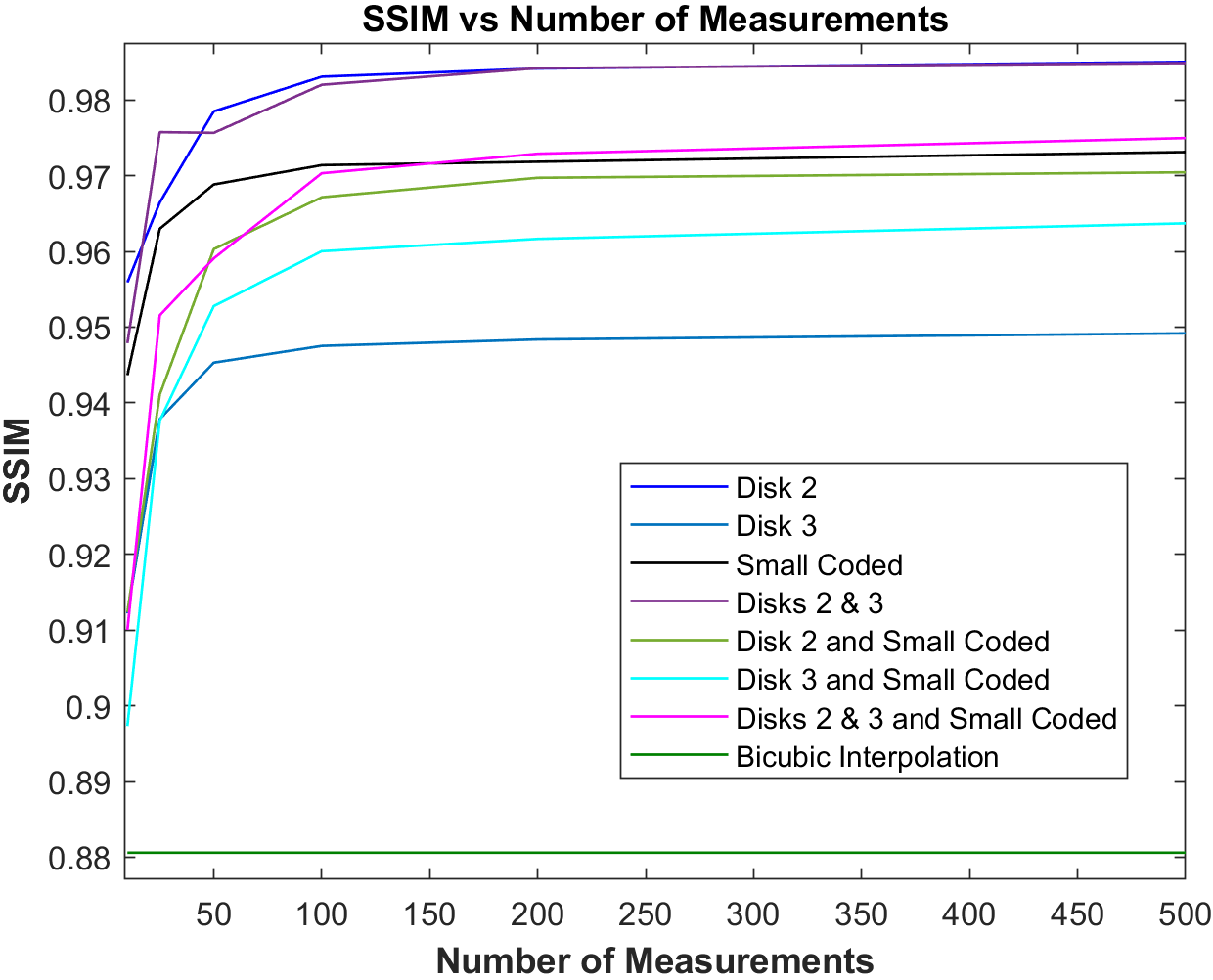}
	\caption{Structural similarity index (SSIM) versus number of random measurements for different kernels, obtained from reconstructions using measurements from the implemented systems.}
	\label{fig:SSIMvsKreal}
\end{figure}

Performance curves for systems using different kernels are shown in Figure \ref{fig:SSIMvsKreal}. The curves roughly track the ordering of the results obtained in simulations, with all kernels and kernel combinations achieving meaningful superresolution. Similar to the simulated cases, the performance improves asymptotically with the number of measurements, yielding only marginal improvements after $50$-$100$ measurements. We note that for a sufficiently fast DMD and sensor array, the total time required to take this many measurements could be on the order of milliseconds.

\begin{figure*}
	\centering
	\includegraphics[scale=0.33]{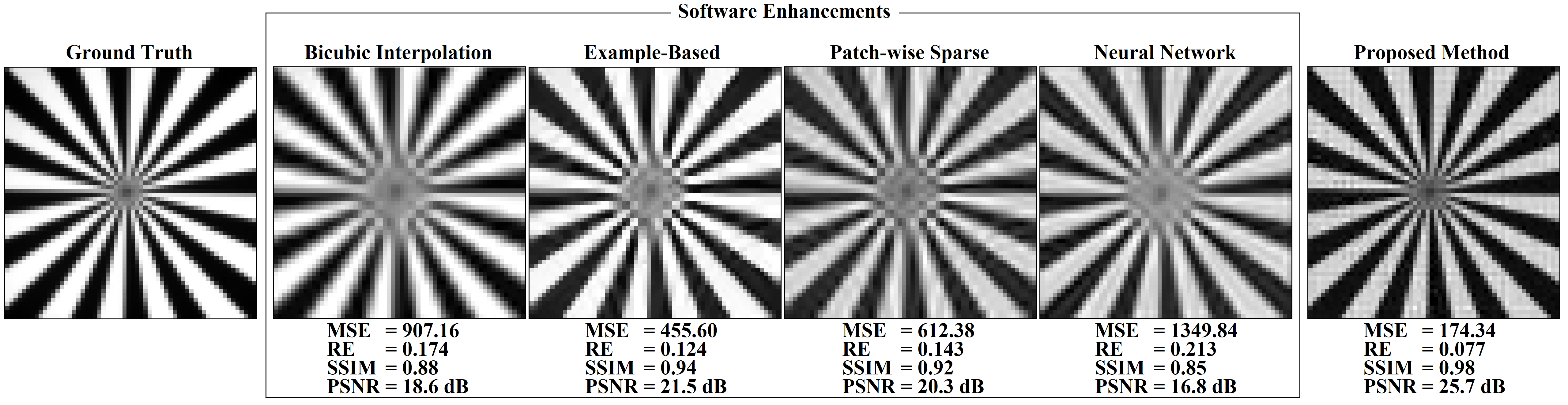}
	\caption{Comparison of the proposed method with several software-based superresolution techniques. The bicubically interpolated low-resolution image was provided as input to all three software-based methods.}
	\label{fig:superrescomparison}
\end{figure*}

In Figure \ref{fig:superrescomparison} we compare a result obtained using the proposed method against results obtained by using several single-image superresolution techniques, including superresolution via patch-wise sparse representation \cite{PatchWiseSparse}, example-based superresolution \cite{ExampleBasedSuperres}, and a superresolving convolutional neural network trained on patches taken from natural images \cite{SRCNN}. This comparison demonstrates the capacity of the system to resolve the image at a higher-level of detail than could be expected with software-based methods alone.


\subsection{Modulation Transfer Function Analysis}

\begin{figure}
	\centering
	\includegraphics[width = 9cm]{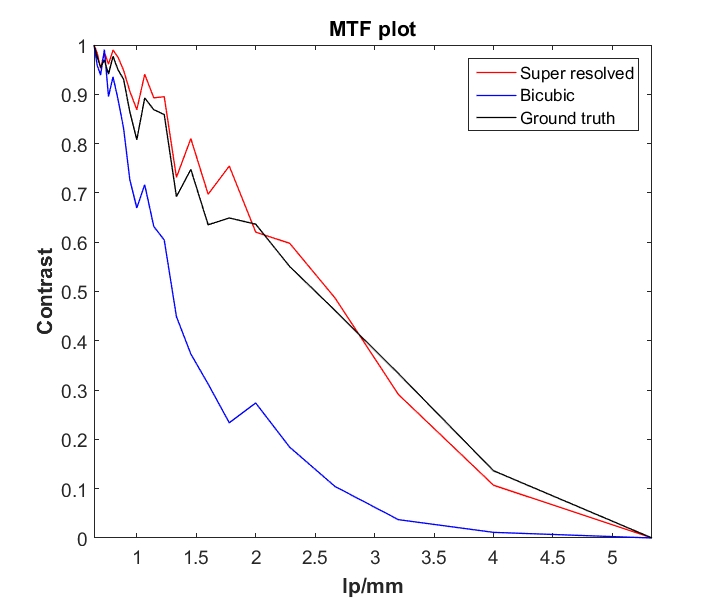}
	\caption{MTF analysis of the super-resolved image, ground truth and bi-cubic interpolated image.}
	\label{fig:mtf}
\end{figure}

To quantify the effect of our method on the spatial frequencies of the output image, the Modulation Transfer Function (MTF){\cite{MTFmeasuring}} was evaluated by calculating the contrast of line pairs in the fan target. Circles with different radii were drawn around the center of the fan target, where the bright/dark line pairs (lp) versus the circle circumference (expressed in lp/mm) decreases as the radius increases. For each circle, the contrast was calculated as $(I_{max}-I_{min} ) / (I_{max}+I_{min} ) $, where $I_{max}$ and $I_{min}$ are the maxima and minima of the bright/dark line pairs matching their locations. Figure \ref{fig:mtf} shows the MTF for each scenario displayed in Figure \ref{fig:exp_res}. The MTF plot clearly verifies the superresolution capability of the proposed method.

\section{Conclusion}

We have demonstrated a novel method for superresolving an imaging system containing an occlusive programmable mask beyond the resolution of both the mask and sensor array. This is distinct from certain compressive imaging systems, such as the single pixel camera, where the resolution of the recovered image is upper-bounded by the resolution of the modulating element. The approach outlined here could be ideal for situations where high-resolution sensor arrays are desired but prohibitively expensive, such as when imaging outside the visible spectrum, in LIDAR, or even for high-speed cameras in the visible range, which usually have resolutions limited by the rate of data transfer off the sensor array.

The resolution enhancement was achieved by inserting a known blur between the mask and sensor and acquiring a series of measurements during which the mask element was randomly modulated. Introducing this blur causes each sensor element to measure a different linear combination of the sub-pixel intensities in the scene, effectively re-casting the superresolution problem into that of inverting a system of linear equations. For a suitable kernel and with sufficiently many random measurements, the resulting linear system can be stably inverted to obtain a superresolved image, and an argument was presented that this will be reliably true for certain kernels in the one-dimensional case.

Under sufficiently many random measurements, the choice of kernel is equivalent (in expectation) to choosing the eigenvalue spectrum of the overall system matrix. The resulting spectrum is then an exact quantitative description of the system's expected performance. We found that a simple defocus gives the imaging system a favorable spectrum, allowing us to superresolve both the mask and sensor by a factor of 4$\times$, although it was found that many different choices of the blur kernel lend the system superresolving capability. Using more complex kernels or multiple kernels did not appear to make a meaningful difference over using a simple blur of a particular size in the case of 4$\times$ superresolution, although could at larger superresolution factors. Building on these results, it is likely possible for a specific application to design an optimal PSF that is more complex than the ones presented here. Extensions to fluttered shutter methods \cite{flutteredshutter} could lead to combined superresolution and motion deblurring. The approach outlined here could also potentially be used to extend the resolution of compressive video systems such as CS-MUVI \cite{csmuvi}.

In this proof-of-concept study we used a simple $\ell_2$-regularized least squares recovery to obtain the superresolved image, but more sophisticated recovery methods are possible. Reasonable models for the structure in natural images, such as low total variation or representation sparsity in a particular dictionary, have been used to regularize multi-image superresolution (\cite{L1Superres}, \cite{SparseMultiframe}), and could be applied here to yield further improvements. It should also be possible to use this system in a compressive manner, obtaining a superresolved image from a much smaller number of measurements.

\bibliographystyle{IEEEtran} 
\bibliography{dmdbib}
\end{document}